\documentclass[12pt]{article}

\usepackage[totalheight = 23cm, totalwidth = 17cm]{geometry}
\usepackage{amssymb,amsmath,amsfonts,amsbsy,graphicx}

\newcommand{\ini}{{\rm (ini)}}
\newcommand{\ad}{{\rm (ad)}}
\newcommand{\nad}{{\rm (nad)}}
\newcommand{\fininf}{{\rm (end)}}
\newcommand{\osc}{{\rm (osc)}}
\newcommand{\fin}{{\rm (fin)}}
\newcommand{\mpl}{m_{\rm Pl}}
\newcommand{\Ggi}{\Gamma_\gamma^{(i)}}
\newcommand{\Gmi}{\Gamma_m^{(i)}}
\newcommand{\Gi}{\Gamma^{(i)}}

\begin{document}

\begin{titlepage}

\begin{center}

\vspace*{-10ex}
\hspace*{\fill}
IFT-UAM/CSIC-08-64, FTUAM 08/20
\\
\hspace*{\fill}
TCC-003-09

\vskip 1.cm

\Huge{Evolution of the curvature perturbation
\\
during and after multi-field inflation}

\vskip 1cm

\large{
 Ki-Young Choi$^{\dag,*}$\footnote{kiyoung.choi@uam.es}
 \hspace{0.2cm}
 Jinn-Ouk Gong$^{\ddag,\S}$\footnote{jgong@lorentz.leidenuniv.nl}
 \hspace{0.2cm}
 Donghui Jeong$^{\P,\star}$\footnote{djeong@astro.as.utexas.edu}
\\
\vspace{0.5cm}
 {\em
 ${}^\dag$Departamento de F\'isica Te\'orica C-XI, Universidad Aut\'onoma de Madrid
 \\
 Cantoblanco, 28049 Madrid, Spain
 \\
 \vspace{0.2cm}
 ${}^*$Instituto de F\'isica Te\'orica UAM/CSIC, Universidad Aut\'onoma de Madrid
 \\
 Cantoblanco, 28049 Madrid, Spain
 \\
 \vspace{0.2cm}
 ${}^\ddag$Department of Physics, University of Wisconsin-Madison
 \\
 1150 University Avenue, Madison, WI 53706-1390, USA
 \\
 \vspace{0.2cm}
 ${}^\S$Instituut-Lorentz for Theoretical Physics, Universiteit Leiden
 \\
 Niels Bohrweg 2, 2333 CA Leiden, The Netherlands\footnote{Present address}
 \\
 \vspace{0.2cm}
 ${}^\P$Department of Astronomy, University of Texas at Austin
 \\
 1 University Station, C1400, Austin, TX 78712, USA
 \\
 \vspace{0.2cm}
 ${}^\star$Texas Cosmology Center, University of Texas at Austin
 \\
 1 University Station, C1400, Austin, TX 78712, USA}
}

\vskip 0.5cm

\today

\vskip 1.2cm

\end{center}

\begin{abstract}

We study the evolution of the curvature perturbation on the super-horizon scales
starting from the inflationary epoch until there remains only a single dynamical
degree of freedom, presureless matter, in the universe. We consider the cosmic
inflation driven by a multiple number of the inflaton fields, which decay into both
radiation and pressureless matter components. We present a complete set of the exact
background and perturbation equations which describe the evolution of the universe
throughout its history. By applying these equations to the simple but reasonable
model of multi-field chaotic inflation, we explicitly show that the total curvature
perturbation is continuously varying because of the non-adiabatic components of the
curvature perturbation generated by the multiple inflaton fields throughout the
whole evolution of the universe. We also provide an useful analytic estimation of
the total as well as matter and radiation curvature perturbations, assuming that
matter is completely decoupled from radiation from the beginning. The resulting
isocurvature perturbation between matter and radiation is at most sub-percent level
when the masses of the inflaton fields are distributed between $10^{-8}\mpl$ and
$10^{-7}\mpl$. We find that this result is robust unless we use non-trivial decay
rates, and that thus, in general, it is hard to obtain large matter-radiation
isocurvature perturbation. Also, by using the $\delta{N}$ formalism, we point out
that the inflationary calculation, especially when involving multiple inflaton
fields, is likely to lose the potentially important post-inflationary evolution
which can modify the resulting curvature perturbation.

\end{abstract}

\end{titlepage}

\setcounter{page}{0}
\newpage
\setcounter{page}{1}

\section{Introduction}

Currently, inflation~\cite{inflation} is the leading candidate of resolving many
cosmological problems such as the horizon, flatness and monopole problems, and of
explaining the initial conditions of the successful hot big bang universe. A certain
amount of quasi-exponential expansion of the universe during inflation out of small
enough region for maintaining causal communication makes the whole observable patch
of the universe homogeneous and isotropic as well as spatially flat. At the same
time, the quantum fluctuations of one or more scalar fields, the inflaton fields,
which dominate the energy density of the universe during inflation are stretched and
become large scale perturbations of spatial curvature. These curvature perturbations
are the seed of the large scale structure and the temperature anisotropy of the
cosmic microwave background (CMB) observed today~\cite{books}. Recent precise
cosmological observations are all consistent with this picture and thus indeed
strongly support the inflation paradigm~\cite{observations,Komatsu:2008hk}.

In spite of the extreme success of the inflation paradigm, the {\em very} inflation
relevant for our own universe is still veiled in mystery. It is because of the
lacking of knowledge on particle physics: the standard model of particle physics
(SM) is one of the most precise theory of physics~\cite{Cottingham:2007zz}, but also
is a limited effective theory of what describes all the interactions in nature at
the very high energy scales relevant for the early universe. For example, the SM
does not explain the neutrino mass~\cite{Bilenky:1987ty}, the baryon
asymmetry~\cite{Riotto:1999yt}, dark matter~\cite{Jungman:1995df}, as well as the
inflaton fields. Fortunately, we have several reasonable and attractive extensions
of the SM such as supersymmetry~\cite{supersymmetry} and extra dimensional
theories~\cite{Rubakov:2001kp}, which presumably will be tested at the Large Hadron
Collider at CERN~\cite{lhc} in the next couple of years. The extended theories of
the SM commonly postulate the existence of a large number of scalar fields other
than the Higgs field which is the sole but yet undiscovered scalar field of the SM.
Therefore, it is very natural to consider inflation models driven by these multiple
number of scalar fields as a plausible realization of inflation in this
context~\cite{Lyth:1998xn}.

The only way to investigate and falsify inflation models is, at best of our current
knowledge, to study the primordial curvature perturbation. It is observed to be
nearly adiabatic and Gaussian, with an almost scale invariant
spectrum~\cite{observations,Komatsu:2008hk}. We naturally expect all of these
properties for single field inflation models. It can be intuitively understood as
follows: as the only degree of freedom during inflation is the single inflaton
field, basically the fluctuations of the inflaton translate into time difference.
This gives rise to perfectly adiabatic perturbations. Also, suppressed self
interaction of the inflaton field required by the observed small deviation from the
scale invariance ensures different Fourier modes of the field fluctuations almost
perfectly independent from each other. This leads to nearly perfect Gaussianity.
Thus, any deviation from perfect adiabaticity or Gaussianity can be a strong
indicator of the existence of more than a single inflaton field. According to the
most recent Wilkinson Microwave Anisotropy Probe (WMAP) 5-year
data~\cite{Komatsu:2008hk}, the observational constraints are such that the
isocurvature perturbation between matter and radiation contributes to the total
density perturbation on the CMB scales at most about 10\%, and the non-linear
parameter $-9<f_\mathrm{NL}^\mathrm{local}<111$ at 95\% confidence level, which
confirms Gaussianity of the primordial perturbation at 0.1\% accuracy.

For inflation models driven by a multiple number of fields, in contrast, there is no
such unique predictions. Given a field trajectory, there are more than one
orthogonal directions into which the field can be `kicked', then the trajectory can
be changed from one to another, i.e. the history of the universe can be entirely
changed. Thus we need to keep tracking the history until different trajectories
coalesce and there remains effectively only one degree of freedom: matter
domination~\cite{deltaN}. Therefore in principle it is never sufficient to consider
only the inflationary phase to make any cosmological predictions, as still a large
number of the inflaton fields survive after inflation and they can subsequently
change the history of the universe after the end of the inflation. For example, let
us consider the curvaton scenario~\cite{curvaton}: in that scenario, one or more
oscillating light scalar fields become dominant component of the energy density of
the universe in the homogeneous radiation background. The observable quantities such
as the power spectrum and its spectral index can be completely different from naive
inflationary predictions, depending on the energy fraction of the curvaton fields at
the moment of their decay. These curvaton fields themselves can be interpreted as
the inflaton fields which have not yet decayed until the radiation dominated era, as
the individual inflaton field can easily satisfy the requirements any curvaton
candidate should do~\cite{Gong:2007gv}.

Therefore, on completely general ground, we should consider the {\em entire}
evolution of the universe until matter domination beyond the inflationary epoch when
we consider multi-field inflation models. The purpose of this paper is to provide a
complete set of background and perturbed equations which describe the universe
starting from multi-field inflation, and to study the evolution of the curvature
perturbation. Especially, we are interested in the isocurvature perturbation between
matter and radiation which can be observationally detected in near future and thus
may be a powerful probe of the early universe. We also discuss the possibility of
large isocurvature perturbation.

The outline of this paper is as follows. In Section~\ref{sec_equations}, we present
the background and perturbed equations of the components which constitute the
universe. They are the inflaton fields, and the decay products of the inflaton
fields: radiation and matter. We write the equations in the spatially flat gauge
which is particularly useful for our purpose. We then give the equations of the
curvature perturbation associated with individual component as well as that of the
total curvature perturbation. In Section~\ref{sec_results}, we present the numerical
results based on the simple chaotic inflation model. We also supply analytic
arguments which helps understand the results. In Section~\ref{sec_discussions} we
discuss the importance of the post-inflationary evolution and the possibility of
generating large isocurvature perturbation. We present our conclusion in Section 5.

\section{Evolution equations}
\label{sec_equations}

In this section we present the evolution equations of both background and perturbed
quantities which constitute the universe. We assume that the universe initially is
dominated by a multiple number of massive scalar fields, $\phi_i$,
$i=1,2,3,~\cdots~N_\mathrm{fields}$, which play the role of the inflaton fields. The
effective potential is taken to be
\begin{equation}
V = V(\phi_1, \phi_2, \cdots) \, .
\end{equation}
Especially, we are interested in the separable potential
\begin{equation}\label{potential}
V = V(\phi_1) + V(\phi_2) + \cdots \equiv V_1 + V_2 + \cdots \, ,
\end{equation}
but nevertheless most of the contents we present in this section should be
applicable to more general classes of potential. Here $\phi_i$ decays into radiation
and matter with the decay rates $\Ggi$ and $\Gmi$ respectively, which are fixed by
underlying physics, e.g. a specific string compactification. We assume for
simplicity that the matter product hardly interacts with radiation even right after
their production. We consider a spatially flat universe, with the metric including
the linear scalar perturbations given by~\cite{Kodama:1985bj}
\begin{equation}\label{eq:matricp}
ds^2 = -(1 + 2A)dt^2 + 2aB_{,i}dtdx^i + a^2 [ (1 - 2\psi)\delta_{ij} + 2E_{,ij} ]
dx^idx^j \, .
\end{equation}
Note that two of the four scalar perturbations $A$, $B$, $\psi$ and $E$ can be
eliminated by specific gauge choice. In this paper, we pick the spatially flat gauge
where the spatial metric is set to be unperturbed, because the perturbation
equations take particularly simple forms in this gauge as we will see soon.

\subsection{Background equations}

First we present the background equations of motion of the inflaton fields $\phi_i$,
including the decay into radiation and matter, as well as those of the radiation and
matter components generated by the decay of the inflaton fields. For
phenomenological description of the decay, we add an additional friction term
$\Gi\dot\phi_i \equiv \left( \Ggi + \Gmi \right)\dot\phi_i$ to the background
equation of motion of $\phi_i$ so that
\begin{equation}\label{BGeomphiitime}
\ddot\phi_i + \left( 3H + \Gi \right) \dot\phi_i + V_{,i} = 0 \, ,
\end{equation}
where $V_{,i} \equiv dV/d\phi_i$. It is more convenient to write the equations in
terms of the number of $e$-folds using
\begin{equation}
dN = H dt \, ,
\end{equation}
from which we find
\begin{align}
\label{dttodN}
\frac{d}{dt} = & H \frac{d}{dN} \, ,
\\
\label{dt2todN2}
\frac{d^2}{dt^2} = & H^2 \frac{d^2}{dN^2} - \frac{\rho + p}{2m_\mathrm{Pl}^2}
\frac{d}{dN} \, ,
\end{align}
with $\mpl \equiv (8\pi{G})^{-1/2}$ being the reduced Planck mass. Here, $\rho$ and
$p$ represent the total energy density and pressure of the system and are given by
\begin{align}
\label{BGtotrhopre}
\rho = & \rho_\gamma + \rho_m + \sum_i \rho_i \, ,
\\
\label{BGtotppre}
p = & p_\gamma + p_m + \sum_i p_i \, ,
\end{align}
respectively, where the energy density and pressure associated with $\phi_i$ are
\begin{align}
\label{BGrhoi}
\rho_i & = \frac{1}{2}\dot\phi_i^2 + V_i = \frac{H^2}{2}{\phi_i'}^2 + V_i \, ,
\\
\label{BGpi}
p_i & = \frac{1}{2}\dot\phi_i^2 - V_i = \frac{H^2}{2}{\phi_i'}^2 - V_i \, ,
\end{align}
with a prime denoting a derivative with respect to $N$. Then
Eqs.~(\ref{BGtotrhopre}) and (\ref{BGtotppre}) become
\begin{align}
\label{BGtotrho}
\rho = & \rho_\gamma + \rho_m + \sum_i \left( \frac{H^2}{2}{\phi_i'}^2 + V_i \right)
\, ,
\\
\label{BGtotp}
p = & \frac{1}{3}\rho_\gamma + \sum_i \left( \frac{H^2}{2}{\phi_i'}^2 - V_i \right)
\, ,
\end{align}
and
\begin{equation}\label{totrhop}
\rho + p = \frac{4}{3}\rho_\gamma + \rho_m + H^2 \sum_i {\phi_i'}^2 \, .
\end{equation}
We can determine the total energy density and pressure by solving the background
evolution equations of the components. Using Eqs.~(\ref{dttodN}) and
(\ref{dt2todN2}), Eq.~(\ref{BGeomphiitime}) becomes
\begin{equation}\label{BGeomphii}
\phi_i'' + \left( 3 + \frac{\Gi}{H} - \frac{\rho + p}{2\mpl^2 H^2} \right) \phi_i' +
\frac{V_{,i}}{H^2} = 0 \, .
\end{equation}

Now we need the evolution equations of the energy densities of radiation and matter
produced by the decay of the inflaton fields $\phi_i$. They can be obtained from the
continuity equation of a certain component $\alpha$, which is derived from the
conservation of energy-momentum tensor
\begin{equation}\label{EMtensorconservation}
T^{\mu\nu}{}_{;\nu} = \sum_\alpha T^{\mu\nu}_\alpha{}_{;\nu} = 0 \, .
\end{equation}
Here a semicolon denotes a covariant derivative and the individual component
$T^{\mu\nu}_\alpha$ needs not be conserved separately, and is given by
\begin{equation}\label{BGrhotime}
\dot\rho_\alpha = -3H(\rho_\alpha + p_\alpha) + Q_\alpha \, ,
\end{equation}
where $Q_\alpha$ denotes the rate of energy transfer related to the $\alpha$
component. In the present case, the only energy transfer is by the decay the
inflation fields with the rate $\Gi$. For the inflaton $\phi_i$, radiation and
matter, they are given by
\begin{align}
Q_i = & -\Gi\dot\phi_i^2 = -H^2\Gi{\phi_i'}^2 \, ,
\\
Q_\gamma = & \sum_i \Ggi\dot\phi_i^2 = H^2 \sum_i \Ggi {\phi_i'}^2 \, ,
\\
Q_m = & \sum_i \Gmi\dot\phi_i^2 = H^2 \sum_i \Gmi {\phi_i'}^2 \, ,
\end{align}
respectively. Then, from Eq.~(\ref{BGrhotime}) the evolution equations are given by
\begin{align}
\label{BGeomrhog}
\rho_\gamma' & = -4\rho_\gamma + H\sum_i \Gamma_\gamma^{(i)}{\phi_i'}^2 \, ,
\\
\label{BGeomrhom}
\rho_m' & = -3\rho_m + H\sum_i \Gamma_m^{(i)}{\phi_i'}^2 \, ,
\end{align}
for radiation and matter, respectively.

Finally, the Hubble parameter $H$ is determined by the Friedmann equation
\begin{equation}\label{Friedmann}
H^2 = \frac{\rho}{3m_\mathrm{Pl}^2} \, ,
\end{equation}
with $\rho$ being given by Eq.~(\ref{BGtotrho}). By taking a derivative with respect
to $N$ and using the continuity equation
\begin{equation}\label{BGcontinuity}
\rho' = -3(\rho + p) \, ,
\end{equation}
we can write the evolution equation of $H$ as
\begin{equation}\label{BGeomH}
H' = -H^{-1} \frac{\rho + p}{2m_\mathrm{Pl}^2} \, .
\end{equation}
Eqs.~(\ref{BGeomphii}), (\ref{BGeomrhog}), (\ref{BGeomrhom}) and (\ref{BGeomH}),
supplemented by Eqs.~(\ref{BGtotrho}), (\ref{BGtotp}) and (\ref{totrhop}), are the
equations we have to solve to find the evolution of the background quantities
$\phi_i$, $\rho_\gamma$, $\rho_m$ and $H$. Note that the total background energy
density $\rho$ evolves according to Eq.~(\ref{BGcontinuity}).

\subsection{Perturbation equations}

In this section, we present the perturbation equations to study the evolution of the
curvature perturbations. As mentioned before we will work in the spatially flat
gauge where the spatial components of the metric perturbation are set to be zero.
The reason why we use the spatially flat gauge is as follows: from the gauge-ready
form of the 00 component of the perturbed Einstein equation in a flat universe,
\begin{equation}\label{gaugeready00Einstein}
3H \left( \dot\psi + HA \right) - \frac{\nabla^2}{a^2} \left[ \psi + H \left( a^2 E
- \dot{a}B \right) \right] = -\frac{\delta\rho}{2\mpl^2} \, ,
\end{equation}
we find that the off-diagonal components of the metric perturbation $B$ and $E$
disappear on the super-horizon scales since they appear only in the spatial gradient
term. Thus, in the spatially flat gauge and on the super-horizon scales, the only
remaining metric perturbation $A$ is directly related to the total density
perturbation
\begin{equation}
\delta\rho = \delta\rho_\gamma + \delta\rho_m + \sum_i \delta\rho_i
\end{equation}
by eliminating the spatial gradient term and $\psi$ from
Eq.~(\ref{gaugeready00Einstein}),
\begin{equation}\label{metricpertA}
A = -\frac{\delta\rho}{2\rho} \, ,
\end{equation}
which makes us free from solving the equation of motion of $A$: we can determine $A$
once we find the total background energy density $\rho$ and its perturbation
$\delta\rho$.

We derive the perturbed equation of motion of the energy density $\rho_\alpha$ by
linearly perturbing the continuity equation, Eq.~(\ref{EMtensorconservation}), we
can find
\begin{equation}\label{eomdeltarho}
\dot{\delta\rho_\alpha} + 3H (\delta\rho_\alpha + \delta p_\alpha) = Q_\alpha A +
\delta Q_\alpha \, ,
\end{equation}
where we take the limit of our interest, i.e. in the flat gauge and on the
super-horizon scales. Now we can use this equation to derive the equation of
$\delta\phi_i$ as well as $\delta\rho_\gamma$ and $\delta\rho_m$. Here one important
point to write the perturbation equations is that, perturbing $\dot\phi_i^2 =
-g^{00}\phi_{i,0}\phi_{i,0}$ would incorporate the metric perturbation
$\delta{g^{00}} = 2A$. That is,
\begin{equation}\label{deltadotphi2}
\delta\left(\dot\phi_i^2\right) = 2\dot\phi_i \left( \dot{\delta\phi_i} -
A\dot\phi_i \right) = 2H^2\phi_i' \left( \delta\phi_i' +
\frac{\delta\rho}{2\rho}\phi_i' \right) \, ,
\end{equation}
where we have used Eq.~(\ref{metricpertA}). We can find the perturbed energy density
and pressure of $\phi_i$ by substituting Eq.~(\ref{BGrhoi}) into
Eq.~(\ref{eomdeltarho}) and using Eq.~(\ref{deltadotphi2}),
\begin{align}
\delta\rho_i = & H^2\phi_i' \left( \delta\phi_i' + \frac{\delta\rho}{2\rho}\phi_i'
\right) + V_{,i}\delta\phi_i \, ,
\\
\delta p_i = & H^2\phi_i' \left( \delta\phi_i' + \frac{\delta\rho}{2\rho}\phi_i'
\right) - V_{,i}\delta\phi_i \, ,
\end{align}
respectively. Thus, using these equations and Eq.~(\ref{BGeomphiitime}), the
background equation of $\phi_i$ with additional assumption that $\Gamma^{(i)}$ is a
constant the equation of motion of $\delta\phi_i$ on the super-horizon scales is
\begin{equation}\label{eq:eomdphi}
\ddot{\delta\phi_i} + 3H\dot{\delta\phi_i} + \Gamma^{(i)}\dot{\delta\phi_i} -
\dot{A}\dot\phi_i + \Gamma^{(i)}A\dot\phi_i + 2AV_{,i} + \sum_j V_{,ij}\delta\phi_j
= 0 \, ,
\end{equation}
or in terms of the number of $e$-folds $N$,
\begin{equation}\label{eomdeltaphii}
\delta\phi_i'' + \left( 3 + \frac{\Gamma^{(i)}}{H} - \frac{\rho +
p}{2m_\mathrm{Pl}^2H^2} \right)\delta\phi_i' - \left\{ \frac{3}{2\rho} \left[
(\delta\rho + \delta p) - (\rho + p) \frac{\delta\rho}{\rho} \right] +
\frac{\Gamma^{(i)}}{H} \frac{\delta\rho}{2\rho} \right\} \phi_i' -
\frac{\delta\rho}{\rho}\frac{V_{,i}}{H^2} + \sum_j\frac{V_{,ij}}{H^2}\delta\phi_j =
0 \, ,
\end{equation}
which gives the evolution of $\delta\phi_i$ coupled to the metric fluctuation. Note
that with the potential given by Eq.~(\ref{potential}), we have
\begin{equation}
V_{,ij} = \frac{d^2}{d\phi_id\phi_j}\sum_kV_k = \frac{d^2V_i}{d\phi_i^2}\delta_{ij}
\, .
\end{equation}
Also we have to solve the perturbation equations of the radiation and matter energy
densities, $\delta\rho_\gamma$ and $\delta\rho_m$. We can find the equations by
using Eqs.~(\ref{eomdeltarho}) and (\ref{deltadotphi2}) as
\begin{align}
\label{eomdeltarhog}
\delta\rho_\gamma' + 4\delta\rho_\gamma - H \sum_i \Gamma_\gamma^{(i)} \left(
{\phi_i'}^2 \frac{\delta\rho}{2\rho} + 2\phi_i'\delta\phi_i' \right) = & 0 \, ,
\\
\label{eomdeltarhom}
\delta\rho_m' + 3\delta\rho_m - H \sum_i \Gamma_m^{(i)} \left( {\phi_i'}^2
\frac{\delta\rho}{2\rho} + 2\phi_i'\delta\phi_i' \right) = & 0 \, ,
\end{align}
for $\delta\rho_\gamma$ and $\delta\rho_m$, respectively.

Finally, we need to supply $\delta\rho$ and $\delta{p}$ which enter
Eqs.~(\ref{eomdeltaphii}), (\ref{eomdeltarhog}) and (\ref{eomdeltarhom}). These are
completely determined by solving these equations. $\delta\rho$ and $\delta{p}$ are
given by
\begin{align}
\label{deltatotrho}
\delta\rho = & H^2 \sum_i \phi_i' \left( \delta\phi_i' +
\frac{\delta\rho}{2\rho}\phi_i' \right) + \sum_i V_{,i}\delta\phi_i +
\delta\rho_\gamma + \delta\rho_m \, ,
\\
\label{deltatotp}
\delta p = & H^2 \sum_i \phi_i' \left( \delta\phi_i' +
\frac{\delta\rho}{2\rho}\phi_i' \right) - \sum_i V_{,i}\delta\phi_i +
\frac{1}{3}\delta\rho_\gamma \, ,
\end{align}
respectively, so that
\begin{equation}\label{deltarhop}
\delta\rho + \delta p = 2H^2 \sum_i \phi_i' \left( \delta\phi_i' +
\frac{\delta\rho}{2\rho}\phi_i' \right) + \frac{4}{3}\delta\rho_\gamma +
\delta\rho_m \, .
\end{equation}
Thus, in the spatially flat gauge on the super-horizon scales\footnote{Keeping
spatial gradient terms, the equation is written as, in the gauge-ready form,
\begin{equation*}
\dot{\delta\rho} + 3H (\delta\rho + \delta{p}) - 3(\rho+p) \left[ \dot\psi -
\frac{\nabla^2}{3} \left( \dot{E} + v \right) \right] = 0 \, ,
\end{equation*}
where $\nabla{v}$ is the perturbed 3-velocity of the fluid.}, the evolution equation
of the total energy density perturbation $\delta\rho$ follows by perturbing the
background continuity equation, Eq.~(\ref{BGcontinuity}), as
\begin{equation}\label{pcontinuity}
\delta\rho' = -3 (\delta\rho + \delta{p}) \, ,
\end{equation}
with $\delta\rho + \delta{p}$ given by Eq.~(\ref{deltarhop}). Therefore, the
perturbation equations to be solved are Eqs.~(\ref{eomdeltaphii}),
(\ref{eomdeltarhog}) and (\ref{eomdeltarhom}). Combined with Eq.~(\ref{pcontinuity})
and the background equations, we can completely determine the perturbed quantities
$\delta\phi_i$, $\delta\rho_\gamma$ and $\delta\rho_m$ as well as $\delta\rho$.

\subsection{Curvature perturbations}

With the calculated background and perturbed quantities, we can write the curvature
perturbations associated with the total energy density and a specific component
$\alpha$, which we denote by $\zeta$ and $\zeta_\alpha$, respectively. The gauge
invariant curvature perturbation $\zeta$ on the uniform density hypersurface is
defined by~\cite{curvatureperturbation}
\begin{equation}\label{zetadefinition}
-\zeta \equiv \psi + H \frac{\delta\rho}{\dot\rho} \, ,
\end{equation}
thus in the flat gauge, $\zeta$ and $\zeta_\alpha$ which is determined by
$\delta\rho_\alpha$ in the same as $\zeta$ are written as
\begin{align}
\label{zeta}
\zeta = & -\frac{\delta\rho}{\rho'} \, ,
\\
\label{zetaalpha}
\zeta_\alpha = & -\frac{\delta\rho_\alpha}{\rho_\alpha'} \, .
\end{align}
From Eqs.~(\ref{zeta}) and (\ref{zetaalpha}), it is straightforward that the total
curvature perturbation $\zeta$ is a weighted sum of all the individual curvature
perturbations $\zeta_\alpha$'s, i.e.
\begin{equation}
\zeta = \sum_\alpha \frac{\rho_\alpha'}{\rho'}\zeta_\alpha \, .
\end{equation}
Now, writing $\zeta_\alpha$'s explicitly, we have
\begin{align}
\label{zetag}
\zeta_\gamma = & \frac{\delta\rho_\gamma}{4\rho_\gamma - H \sum_i \Ggi {\phi_i'}^2}
\, ,
\\
\label{zetam}
\zeta_m = & \frac{\delta\rho_m}{3\rho_m - H \sum_i \Gmi {\phi_i'}^2} \, ,
\\
\label{zetai}
\zeta_i = & \frac{\phi_i' \left[ \delta\phi_i' + \phi_i' \delta\rho/(2\rho) \right]
+ V_{,i}\delta\phi_i/H^2}{\left( 3 + \Gi/H \right){\phi_i'}^2} \, ,
\end{align}
where we have used
\begin{align}
\rho_i' = & -H^2 \left( 3 + \frac{\Gamma^{(i)}}{H} \right) {\phi_i'}^2 \, ,
\\
\delta\rho_i = & H^2 \left[ \phi_i' \left( \delta\phi_i' + \phi_i'
\frac{\delta\rho}{2\rho} \right) + \frac{V_{,i}}{H^2}\delta\phi_i \right] \, .
\end{align}
Thus, Eqs.~(\ref{zetag}), (\ref{zetam}), (\ref{zetai}), and the total curvature
perturbation
\begin{align}\label{zetatot}
\zeta = & -\frac{\delta\rho_\gamma + \delta\rho_m + \sum_i
\delta\rho_i}{\rho_\gamma' + \rho_m' + \sum_i \rho_i'}
\nonumber\\
= & \frac{\delta\rho_\gamma + \delta\rho_m + H^2 \sum_i \left\{ \phi_i' \left[
\delta\phi_i' + \phi_i'\delta\rho/(2\rho) \right] + V_{,i}\delta\phi_i/H^2
\right\}}{4\rho_\gamma + 3\rho_m + 3H^2 \sum_j {\phi_j'}^2} \, ,
\end{align}
as well as the isocurvature perturbation between any two
components~\cite{isocurvature}
\begin{equation}\label{isocurvatureS}
\mathcal{S}_{\alpha\beta} = 3 \left( \zeta_\alpha - \zeta_\beta \right) = -3 \left(
\frac{\delta\rho_\alpha}{\rho_\alpha'} - \frac{\delta\rho_\beta}{\rho_\beta'}
\right) \, ,
\end{equation}
are completely determined by solving the background equations,
Eqs.~(\ref{BGeomphii}), (\ref{BGeomrhog}), (\ref{BGeomrhom}) and (\ref{BGeomH}), and
the perturbation equations, Eqs.~(\ref{eomdeltaphii}), (\ref{eomdeltarhog}) and
(\ref{eomdeltarhom}). These solutions describe the evolution of the curvature
perturbations given by Eqs.~(\ref{zetag}), (\ref{zetam}), (\ref{zetai}) and
(\ref{zetatot}) in the universe composed of the inflaton fields $\phi_i$ and their
decay products, radiation and matter. Note that although $\zeta_\gamma$, $\zeta_m$
and $\zeta_i$ may meet singularities when the denominators become zero, that of
$\zeta$ remains always finite and thus $\zeta$ is well defined throughout the
evolution of the universe.

\section{Evolution of the curvature perturbations}
\label{sec_results}

In this section, we use the equations we derived in the previous section and
explicitly study a universe initially filled with a multiple number of the inflaton
fields until pressureless matter, which is produced by the decay of the inflatons,
becomes the most dominant component. For simplicity, we consider the case of
multiple chaotic inflation with the potential
\begin{equation}\label{multichaotic}
V = \sum_i V_i = \sum_i \frac{1}{2} m_i^2\phi_i^2 \, .
\end{equation}
We first solve the equations given in the previous sections numerically in
Section~\ref{numerical}, then estimate the resulting curvature perturbation along
with the isocurvature perturbation analytically in Section~\ref{analytic}.

\subsection{Numerical results}
\label{numerical}

In this section, we will present the details of the numerical implementation we
adopt in order to solve the equations we presented in the previous section.

\subsubsection{Parameters}

\begin{figure}
\centering
\rotatebox{90}{%
  \includegraphics[width=12cm]{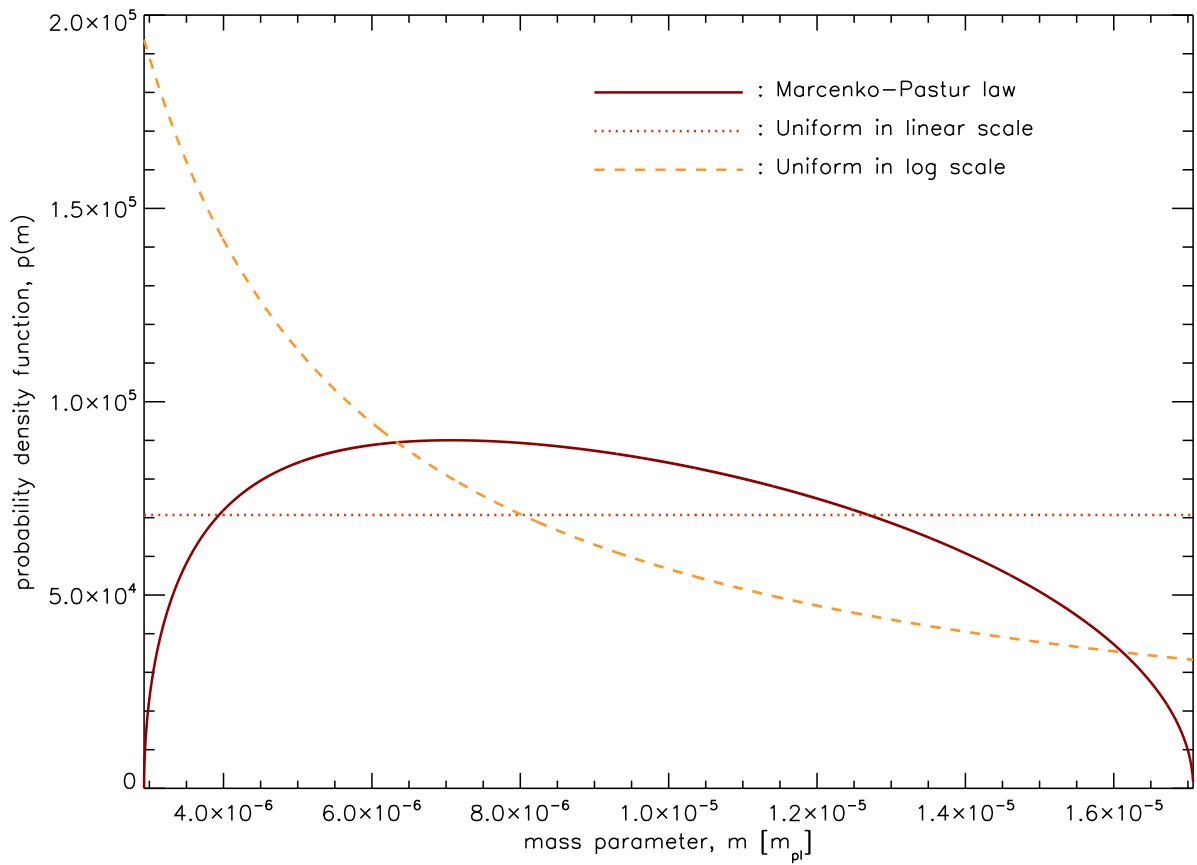}
}%
\caption{%
The probability density functions, $p(m)$, of three different mass distributions:
Mar{\v c}enko-Pastur distribution, uniform distribution in linear scale, and uniform
distribution in log scale. We use $\bar{m}=10^{-5}\mpl$ and $\beta=0.5$ for the
Mar{\v c}enko-Pastur distribution , and other distributions are uniform between
$m_1$ and $m_{N_\mathrm{fields}}$ given by the Mar{\v c}enko-Pastur distribution in
linear and log scales, respectively.
}%
\label{fig0}
\end{figure}

We consider the case of $N_\mathrm{fields} = 1000$ inflaton fields whose masses are
assigned by three different distributions:
\begin{itemize}

\item[(i)] \textit{Mar{\v c}enko-Pastur distribution}\\
The probability density function of the Mar{\v c}enko-Pastur
distribution~\cite{MPdistribution,Easther:2005zr} is
\begin{equation}\label{MPdist}
p(m^2)=
\frac{
    \sqrt{
        \left[\bar m^2(1+\sqrt\beta)^2-m^2\right]
        \left[m^2-\bar m^2(1-\sqrt\beta)^2\right]
    }
}
{2\pi\beta \bar m^2 m^2},
\end{equation}
for $\bar{m}^2(1-\sqrt\beta)^2<m^2<\bar{m}^2(1+\sqrt\beta)^2$, and zero otherwise.
Here, $\bar m$ is the average of the mass squared, i.e. $\langle m^2\rangle=\bar
m^2$ and $\beta$ is a model dependent parameter which quantifies the broadness of
the distribution. In the calculation, we set $\bar m=10^{-5}\mpl$ and $\beta=0.5$,
and the inflaton field masses are distributed between $m_1=1.69959\times 10^{-5}
\mpl$ and $m_{N_\mathrm{fields}}=2.97069\times 10^{-6}\mpl$.

\item[(ii)] \textit{Uniform distribution in linear scale}\\
We distribute the mass uniformly in linear scale between $m_1$ and
$m_{N_\mathrm{fields}}$ of case (i). That is, the probability density function is
\begin{equation}
p(m) = \frac{1}{m_1-m_{N_\mathrm{fields}}}
\end{equation}
when $m_{N_\mathrm{fields}}<m<m_1$, and otherwise zero.

\item[(iii)] \textit{Uniform distribution in log scale}\\
Finally, we distribute the mass uniformly in log scale between $m_1$ and
$m_{N_\mathrm{fields}}$ of case (i). In this case, the probability density function
is
\begin{equation}
p\left(\log m\right) = \frac{1}{\log m_1 - \log m_{N_\mathrm{fields}}}
\end{equation}
when $m_{N_\mathrm{fields}}<m<m_1$, and otherwise zero.

\end{itemize}
We show the probability distribution function for each distribution in
Figure~\ref{fig0}. Note that we do not randomly distribute the mass. Instead, we
invert the cumulative probability distribution function $F(x)=\int_0^{x} p(x')dx'$,
and assign the mass of $i$-th field by
\begin{equation}
m_{i} = F^{-1}\left(\frac{N_{\mathrm{fields}}-i+0.5}{N_\mathrm{fields}}\right) \, ,
\end{equation}
with $i = 1,2,3,\cdots N_\mathrm{fields}$, so that $\phi_1$ and
$\phi_{N_\mathrm{fields}}$ become the heaviest and the lightest fields,
respectively.

Since we do not know the exact decay rate, we use $\Gamma^{(i)}$ proportional to the
cubic power of the mass of the corresponding inflaton field. In addition, in order
for the matter component to remain sub-dominant since the big bang nucleosynthesis
(BBN) until matter-radiation equality, the decay rate to radiation has to be at
least $10^6$ larger than that to matter~\cite{Gupta:2003jc}. Therefore, we choose
\begin{align}
\Gamma_\gamma^{(i)} = & C \frac{m_i^3}{\mpl^2} \, ,
\\
\Gamma_m^{(i)} = & 10^{-6} \Gamma_\gamma^{(i)} \, ,
\end{align}
where we require $C \ge 10^6$: this condition comes from that the minimum decay rate
we can think of is that of gravitational decay, where the decay rate is proportional
to the cube of the mass, thus $\Gamma^{(i)}_m \ge m_i^3/\mpl^2$\footnote{This also
helps computationally, since too small $C$ requires infeasible amount of computing
time as the decay of the inflaton fields happens very slowly.}. Note that by this
difference we are assuming that matter component is already non-relativistic when it
is produced by the decay of the inflatons.

\begin{figure}
\centering
\rotatebox{90}{%
  \includegraphics[width=12cm]{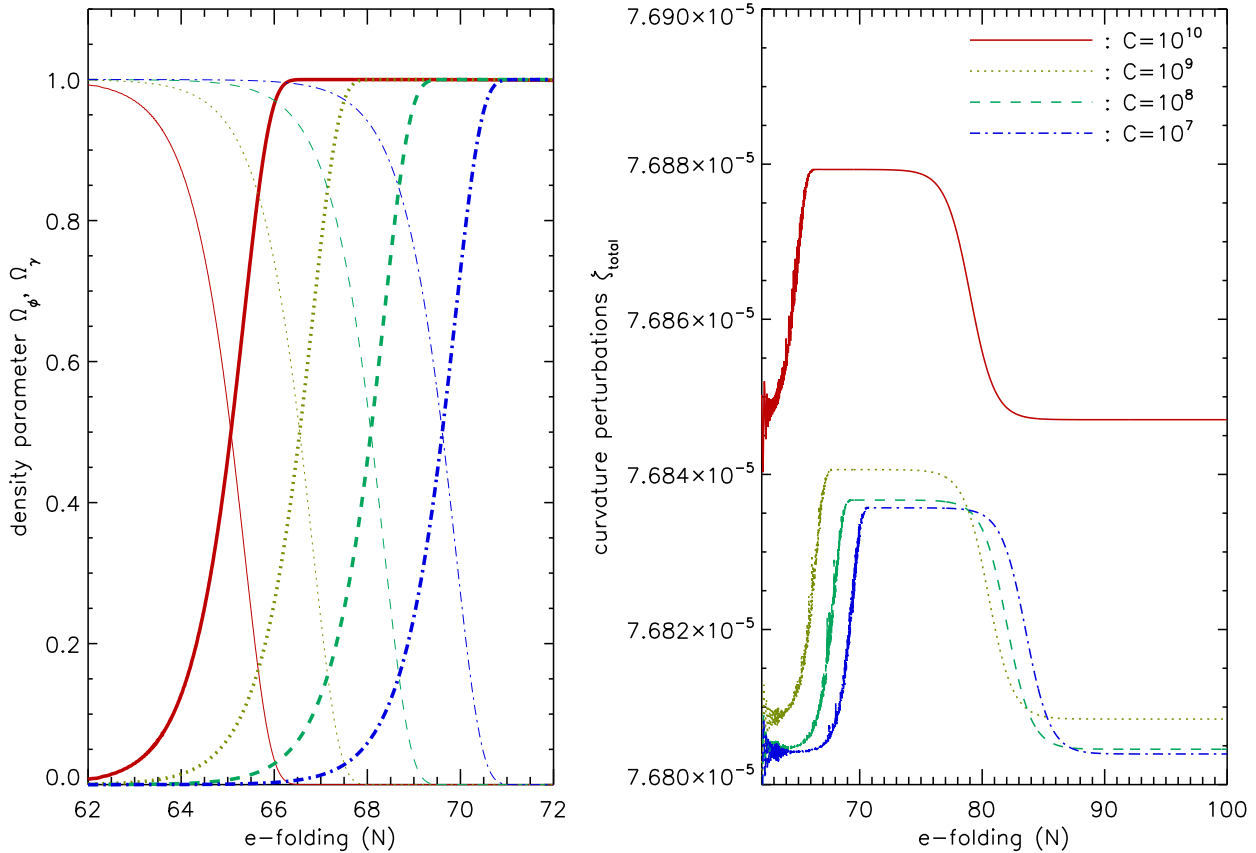}
}%
\caption{%
Comparison of density parameter (left) and the total curvature perturbation (right)
for four different decay rates: $C=10^{10}$ (red, solid line), $10^9$ (olive, dotted
line), $10^8$ (dark green, dashed line), and $10^7$ (blue, dot-dashed line), where
$\Gamma^{(i)}_\gamma=Cm_i^3/\mpl^2$. We use $100$ inflaton fields whose masses
follow the Mar{\v c}enko-Pastur distribution with $\bar{m}=10^{-6}\mpl$ and
$\beta=0.5$. In the left panel, thick lines show $\Omega_\gamma$ and thin ones
$\Omega_\phi$ for each case. As large $C$ increases the decay rate, the inflatons
decay faster then. In the right panel, all lines show the corresponding total
curvature perturbations. The resulting curvature perturbations remain almost same in
the sub-percent level when $C<C_\mathrm{crit}\equiv {m_{100}\mpl^2}/{m_{1}^3}$.
}%
\label{fig0.5}
\end{figure}

Therefore, we can choose the constant $C$ arbitrarily while the highest decay rate,
$\Gamma^{(1)}$, remains much smaller than the lightest mass,
$m_{N_\mathrm{fields}}$, and we use $C=10^{6}$ in this paper: in the current case,
we require
\begin{equation}
\Gamma^{(1)} = C \frac{m_{1}^3}{\mpl^2} \ll m_{N_\mathrm{fields}} \, ,
\end{equation}
which leads to
\begin{equation}
C \ll \frac{m_{N_\mathrm{fields}}\mpl^2}{m_{1}^3} \equiv C_\mathrm{crit} =
6.05097\times 10^{8} \, .
\end{equation}
As long as the inflaton fields decay while they are all rapidly oscillating, the
final curvature perturbation is almost the same irrespective of $C$, i.e. as long as
$C < C_\mathrm{crit}$. Then, how small should $C$ be compare to $C_\mathrm{crit}$?
We have tested that for $N_\mathrm{fields}=100$ case with the Mar{\v c}enko-Pastur
mass distribution of $\bar{m}=10^{-6}\mpl$ and $\beta=0.5$. In this case,
$C_\mathrm{crit}={m_{100}\mpl^2}/{m_{1}^3}\sim 10^{10}$. By comparing $C=10^7$,
$10^8$, $10^9$ and $10^{10}$, we confirm that the resulting curvature perturbation
remains almost same in the sub-percent level while $C<C_\mathrm{crit}$. Especially,
when $C<10^{8}\sim10^{-2}C_\mathrm{crit}$, $\zeta$ is almost exactly the same.
Therefore, for $N_\mathrm{fields}=1000$, $\bar{m}=10^{-5}$ cases, our choice of
$C=10^{6}\sim10^{-2}C_\mathrm{crit}$ is safe. The differences basically arises
because the inflatons decay slowly for lower value of $C$. We show the results in
Figure~\ref{fig0.5}.

\subsubsection{Initial conditions}

In this section, we denote the initial value of a quantity by a subscript `(ini)'.
For simplicity, we assume that the initial field value is the same for every field,
and is set to make the number of the $e$-folds at the end of inflation equal to
about $60$ with the estimation\footnote{See, however, Section~\ref{subsec_deltaN}.}
\begin{equation}
N = \sum_i \frac{\phi_{i\ini}^2}{4\mpl^2} \, .
\end{equation}
We assume that the slow-roll condition is satisfied initially, and set
initial field velocity as
\begin{equation}
\phi_{i\ini}' = -\frac{V_{,i}}{3H_\ini^2} = -\frac{m_i^2\phi_{i\ini}}{3H_\ini^2} \, .
\end{equation}
Here initial Hubble constant, $H_\ini$, is set to satisfy the Friedmann equation
\begin{equation}
3H_\ini^2 = \rho_{m\ini} + \rho_{\gamma\ini} + \frac{H_\ini^2}{2} \sum_i {\phi_{i\ini}'}^2
+ \frac{1}{2}\sum_i m_i^2\phi_{i\ini}^2 \, .
\end{equation}
As any pre-existing matter or radiation energy density will be exponentially diluted
away, we set them initially zero, i.e.
\begin{equation}
\rho_{m\ini} = \rho_{\gamma\ini} = 0 \, .
\end{equation}
The above equations completely specify the initial conditions for the background
evolution. Now, we set the initial conditions for the perturbations.

To compare the numerical results with the analytic estimates, which are performed in
the spatially flat gauge and in the longitudinal gauge (see Section~\ref{analytic})
respectively, we match the field fluctuations in these two gauges by using a gauge
invariant variable~\cite{sasakimukhanov}
\begin{equation}
Q_i = \delta\phi_i + \phi_i'\psi \, ,
\end{equation}
so that
\begin{equation}
\delta\phi_i^{(F)} = \phi_i' \left[ \frac{\delta\phi_i^{(L)}}{\phi_i'} +\psi^{(L)}
\right] \, ,
\end{equation}
where the superscripts `($F$)' and `($L$)' denote the spatially flat and the
longitudinal gauges, respectively. Note that the terms in the square brackets on the
right hand side constitute the curvature perturbation $\zeta_i$, which makes sense
since in the spatially flat gauge simply $\psi^{(F)} = 0$. The initial metric
perturbation in the longitudinal gauge $\psi_\ini^{(L)}$, which is denoted as $\Phi$
in Section~\ref{analytic}, is obtained from Eqs.~(\ref{SRPhisol}) and
(\ref{SRdeltaphiisol}) along with Eqs.~(\ref{coeff_C1}) and (\ref{coeff_di}). For
computational convenience, $\delta\phi^{(L)}_{i\ini}$ is set by the typical value
\begin{equation}\label{eq:dphiL}
\delta\phi^{(L)}_{i\ini} = \frac{H_\ini}{2\pi} \, .
\end{equation}
We set the velocity of initial perturbation by imposing the slow-roll condition:
under the slow-roll approximation, Eq.~(\ref{eq:eomdphi}) and $A=\delta\rho/(2\rho)$
are approximated as
\begin{align}
& 3H\delta\dot\phi_i+\Gamma^{(i)}\delta\dot\phi_i +\Gamma^{(i)}A\dot\phi_i
+2Am_i^2\phi_i + m_i^2\delta\phi_i = 0 \, ,
\\
& A = \frac{\sum_i m_i^2\phi_i\delta\phi_i} {2\rho \left[ 1-\sum_j
H^2\phi_j'^2/(2\rho) \right]} \, ,
\end{align}
respectively. Therefore, we set
\begin{equation}
\delta\phi_{i\ini}' = \frac{-1}{1+\Gamma^{(i)}/(3H_\ini)} \left[
\frac{m_i^2\delta\phi_{i\ini}}{3H_\ini^2}+
\frac{A_\ini}{3H_\ini^2}\left(2m_i^2\phi_{i\ini} +
\Gamma^{(i)}H_\ini\phi_\ini'\right) \right] \, .
\end{equation}
As we assume that there is no matter or radiation energy density to be perturbed,
again simply
\begin{equation}
\delta\rho_{m\ini} = \delta\rho_{\gamma\ini} = 0 \, .
\end{equation}

\subsubsection{Evolution}

We solve the system of the coupled ordinary equations by using the eighth order
adaptive Runge-Kutta method~\cite{nr:2008}. For numerical efficiency, after the end
of inflation, we drop a field from the dynamics if the energy density of that field
becomes smaller than $10^{-5}$ of radiation {\em and} matter energy densities. That
is, $\phi_i$ is dropped from the dynamical equations when {\em both}
\begin{align}
\Omega_i < & 10^{-5} \Omega_\gamma \, ,
\\
\Omega_i < & 10^{-5} \Omega_m \, ,
\end{align}
are satisfied. We have also checked different criteria, but the results hardly
change. This drop-out is valid in our example since heavier fields have smaller
energy densities during oscillation and decay earlier, so that they do not dominate
the energy density later again.

\subsubsection{Results}

\begin{figure}
\centering
\rotatebox{90}{%
  \includegraphics[width=12cm]{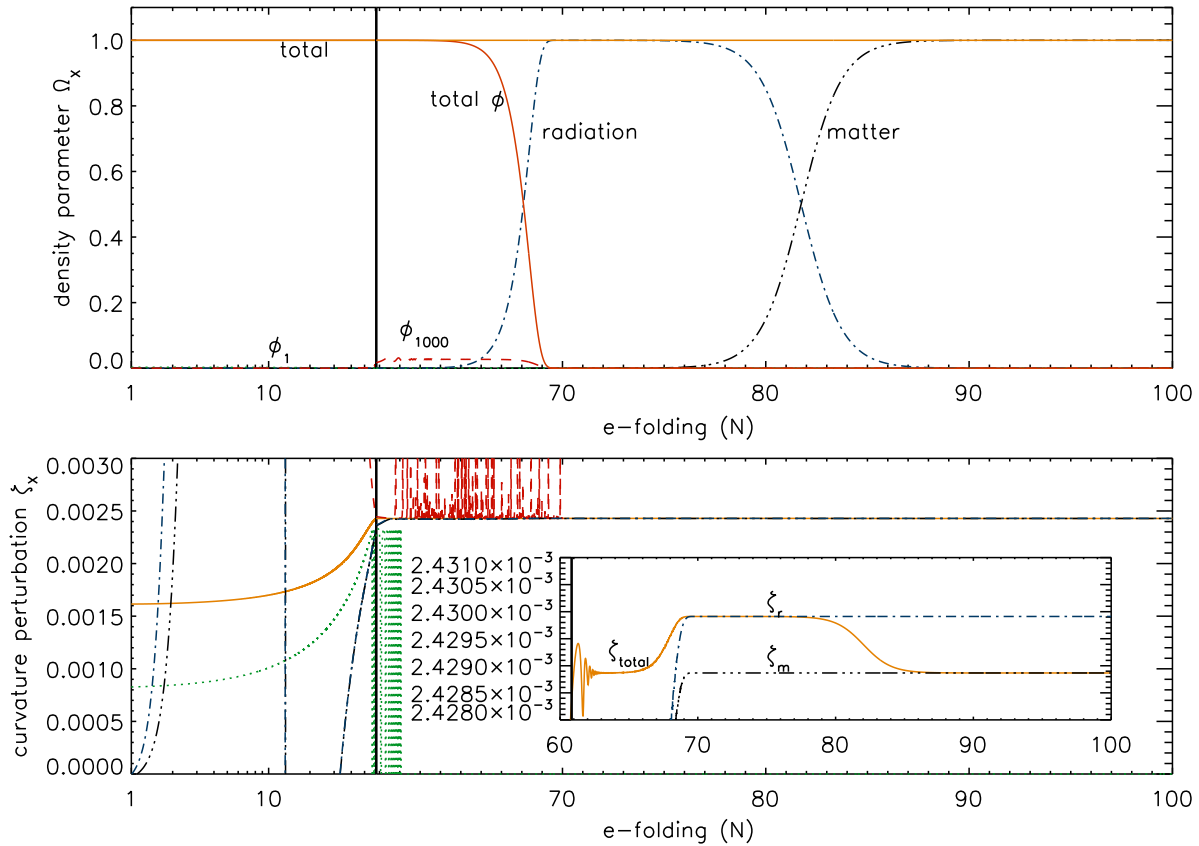}
}%
\caption{%
Evolution of the density parameter (top) and the super-horizon mode curvature
perturbation (bottom) of the universe filled with radiation, matter and $1000$
inflaton fields, whose mass distribution follows the Mar{\v c}enko-Pastur
distribution with $\bar{m}=10^{-5}\mpl$ and $\beta=0.5$. We show the density
parameters and the curvature perturbations associated with the heaviest and the
lightest inflaton fields (green, dotted and red, dashed lines, respectively),
radiation (dark blue, dot-dashed line) and matter (black, dots-dashed line) as well
as the total ones (solid, orange line). We also denote the end of the inflation as a
thick vertical line. In the bottom panel, we highlight the late time evolution of
the total curvature perturbation along with matter and radiation ones. For
multi-field inflation, the curvature perturbation changes on the super-horizon
scales during and after inflation, and there exists a slight amount of the
matter-radiation isocurvature perturbation.
}%
\label{case1}
\end{figure}

\begin{figure}
\centering
\rotatebox{90}{%
  \includegraphics[width=12cm]{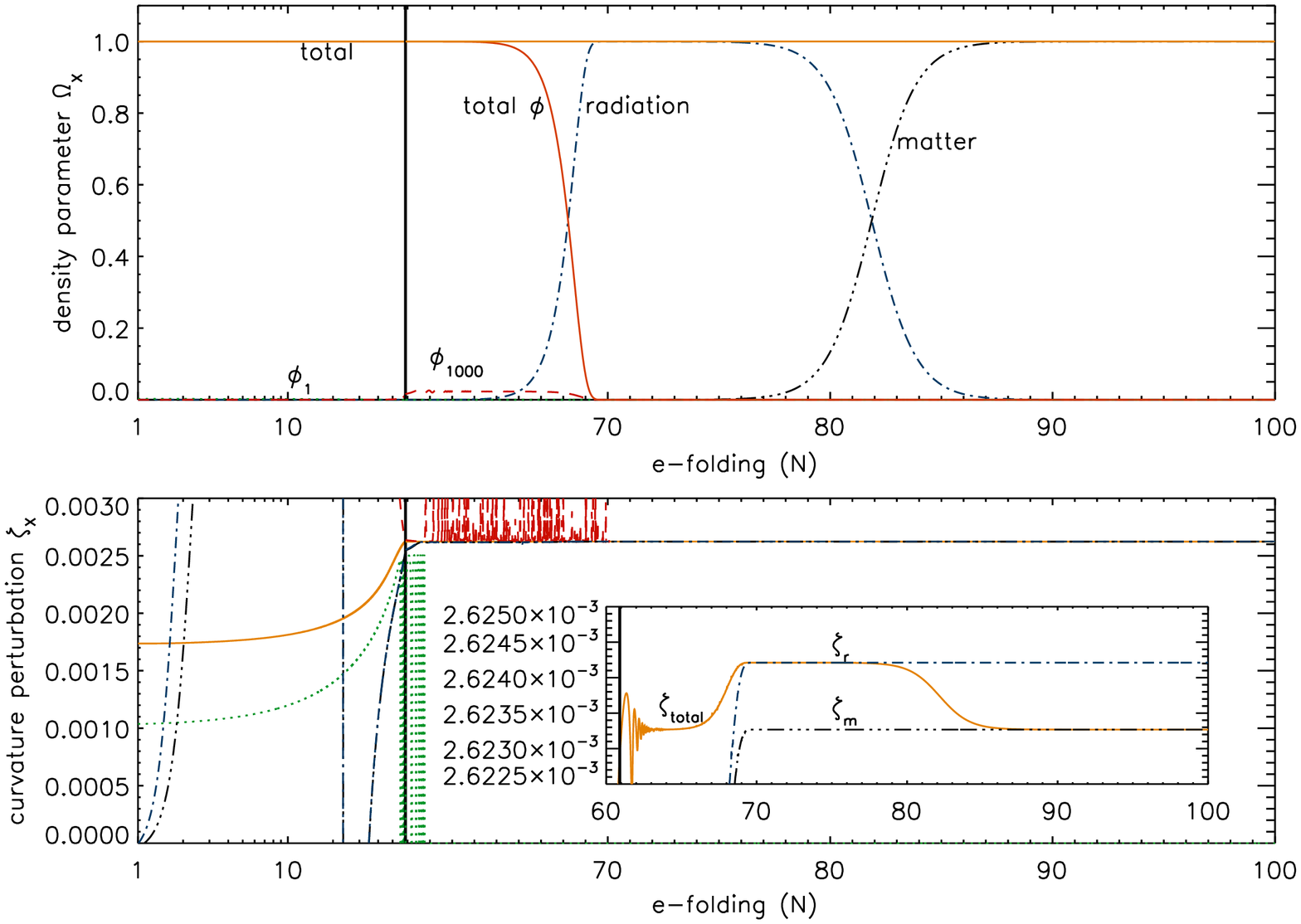}
}%
\caption{%
The same as Figure~\ref{case1}, except that the masses of the inflaton fields are
distributed according to the uniform distribution in linear scale between
$m_1=1.69959\times 10^{-5} \mpl$ and $m_{1000}=2.97069\times 10^{-6} \mpl$.
}%
\label{case2}
\end{figure}

\begin{figure}
\centering
\rotatebox{90}{%
  \includegraphics[width=12cm]{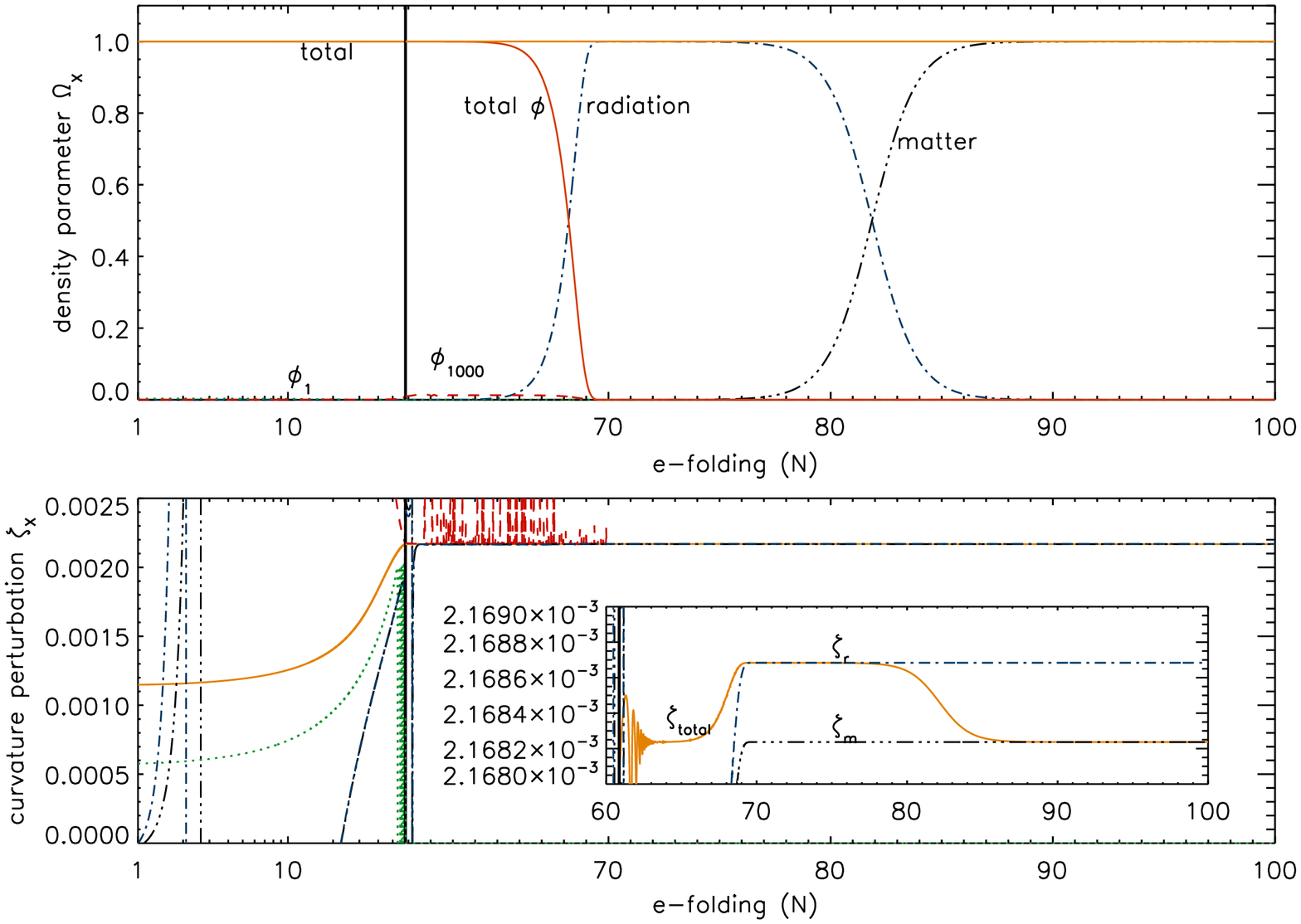}
}%
\caption{%
The same as Figure~\ref{case1}, but the masses of the inflaton fields are
distributed according to the uniform distribution in log scale between
$m_1=1.69959\times 10^{-5} \mpl$ and $m_{1000}=2.97069\times 10^{-6} \mpl$.
}%
\label{case3}
\end{figure}

Figures~\ref{case1}, \ref{case2} and \ref{case3} show the results of our numerical
calculations. In these figures, we show the evolution of the density parameters (top
panel) and the curvature perturbations (bottom panel) of the heaviest (green, dotted
line) and the lightest (red, dashed line) inflaton fields, the radiation (dark blue,
dot-dashed line), and the matter (black, dots-dashed line). The total curvature
perturbation and the density parameter are also shown with orange solid line. The
end of inflation, when the slow-roll parameter
\begin{equation}\label{SRepsilon}
\epsilon \equiv -\frac{\dot{H}}{H^2} = -\frac{H'}{H}
\end{equation}
becomes unity, is also shown as a thick vertical line around $N=60$. In order to
analyze better the evolution of the curvature perturbation after inflation, we
magnify the curvature perturbation between $60 \leq N \leq 100$ in the bottom panel.

After inflation ends, the decay products of the inflatons start dominating the
universe. Since we set the decay rate to the radiation $10^{6}$ larger than that to
the pressureless matter, radiation dominate first and then matter domination comes
next. We set the end of the reheating period and the beginning of the radiation
dominated era when the radiation energy density occupies more than half of the total
one, i.e. $\Omega_\gamma \geq 0.5$. The temperature of the universe at this time, or
the ``reheating temperature'' $T_\mathrm{RH}$, is calculated from~\cite{Kolb:1990vq}
\begin{equation}
\rho_\gamma = \frac{\pi^2}{30}g_* T^4 \, ,
\end{equation}
where $g_*$ represents the effective degrees of freedom of radiation, which is
related to the total number of particle species. For definiteness, we set
$g_*=10^2$, and calculate the reheating temperature for each cases. The results are
summarised in Table~\ref{table:bg_summary}.

\begin{table}[h]
\centering
\begin{tabular}{c|c|c|c}
\hline \hline
   & Mar{\v c}enko-Pastur law & uniform linear & uniform log\\
\hline
$N_\mathrm{end}$         & 60.8433& 60.9048   &60.8669\\
\hline
$N_\mathrm{RH}$   &68.0806 &68.2138&68.2071 \\
\hline
$T_\mathrm{RH}/10^{12}\mathrm{GeV}$
&5.91741 & 5.38037 & 5.07352
\\
\hline \hline
\end{tabular}
\caption{%
Number of $e$-folds at the end of inflation and at the end of the reheating
period, and the reheating temperature.
}
\label{table:bg_summary}
\end{table}

In the bottom panels of the Figures~\ref{case1}-\ref{case3} in this section, we find
that the curvature perturbation of ${\phi_{1}}$ (heaviest field) and ${\phi_{1000}}$
(lightest field) show spikes near the end of inflation. These spikes are due to the
oscillation of the fields, as $\phi_i'$ becomes very small near the maximum
amplitude of the oscillation. In addition to that, the matter and radiation
curvature perturbations, denoted by $\zeta_m$ and $\zeta_\gamma$ respectively,
change drastically during inflation. We can attribute these features to the fact
that the curvature perturbation of each component is ill-defined when the
denominator $\rho_\alpha'$ occasionally becomes zero. However, total curvature
perturbation $\zeta_\mathrm{total}$ is very well defined throughout the whole
evolution.

We find that, for inflation driven by multiple scalar fields, the total curvature
perturbation constantly changes on the super-horizon scale contrast to the single
filed case, even after inflation. The total curvature perturbation increases during
inflation, and even oscillates when the inflaton fields are in the oscillatory
phase. In the radiation and matter dominated epochs, it remains constant following
the curvature perturbation of the most dominant component, but it changes again
during the transition phase around the moment of matter-radiation equality. We also
find that while multiple inflaton fields can generate the isocurvature perturbation
between matter and radiation, its amount is very tiny compared with the total
curvature perturbation. For all of the three mass distributions, the isocurvature
perturbation is only about $0.1\%$ of the final total curvature perturbation. The
results are summarized in Table~\ref{table:summary}.

\begin{table}[h]
\centering
\begin{tabular}{c|c|c|c|c}
\hline \hline
 \multicolumn{2}{r|}{}  & Mar{\v c}enko-Pastur law & uniform linear & uniform log\\
\hline
$\zeta_m$ & numerical& $2.42887\times 10^{-3}$ & $2.62327\times 10^{-3}$ & $2.16824\times 10^{-3}$\\
\cline{2-5}
          & analytic & $2.47077\times 10^{-3}$ & $2.66871\times 10^{-3}$& $2.21415\times 10^{-3}$\\
\hline
$\zeta_\gamma$ & numerical &$2.42991\times 10^{-3}$ & $2.62421\times 10^{-3}$& $2.16868\times 10^{-3}$\\
\cline{2-5}
          & analytic  & $2.47247\times 10^{-3}$ & $2.67120\times 10^{-3}$& $2.21625\times 10^{-3}$\\
\hline
$S_{mr}$  & numerical & $-3.13389\times 10^{-6}$ & $-2.83071\times 10^{-6}$& $-1.34026\times 10^{-6}$\\
\cline{2-5}
          & analytic & $-5.10323\times 10^{-6}$ & $-7.46318\times 10^{-6}$& $-6.29211\times 10^{-6}$\\
\hline \hline
\end{tabular}
\caption{ Comparison between the numerical calculations in Section~\ref{numerical}
and the analytic estimations in Section~\ref{analytic}. We compare the final
curvature perturbation of matter and radiation, along with the isocurvature
perturbation between them. Analytic estimation provides reasonably good agreement:
it predict the curvature perturbations about $2\%$ accuracy, and the isocurvature
perturbation within a factor of $\mathcal{O}(1)$.}
\label{table:summary}
\end{table}

\subsection{Analytic estimates}
\label{analytic}

As we have seen in the previous section the curvature perturbation $\zeta$ is not
conserved even on the super-horizon scales during and after inflation in the
presence of a multiple number of the inflaton fields. To the best of our knowledge,
however, there is no way to describe the evolution of $\zeta$ {\em throughout} the
entire regime of the slow-roll inflation, the phase of the inflaton oscillation and
the subsequent radiation and matter dominated epochs. The difficulty arises because
the inflaton fields are not simple barotropic system which satisfies $p_\phi =
p_\phi(\rho_\phi)$, i.e. in general the equation of state of the inflaton field is a
non-trivial function of $\phi$ and cannot be written as a simple function of
$\rho_\phi$ like radiation ($p_\gamma = \rho_\gamma/3$) or matter ($p_m = 0$).
Nevertheless, we do find the evolution of $\zeta$ during each phase in the
literatures. In this section, by combining those results together we try to
understand the behavior of $\zeta$ presented in the previous section.

\subsubsection{During slow-roll inflation}

Ref.~\cite{Polarski:1994rz} provides a set of general solutions on the super-horizon
scales under the slow-roll approximation with a separable potential by sum such as
Eq.~(\ref{potential}). With this potential of the form $V = \sum_iV_i(\phi_i)$, the
general growing\footnote{The `decaying' adiabatic solutions are written as $\Phi =
C_2 H/a$ and $\delta\phi_i/\dot\phi = -C_2/a$, respectively, with $C_2$ some
constant.  But as can be read from these solutions, their contributions to $\zeta_i$
exactly cancel each other and we can completely neglect the decaying
solutions~\cite{Polarski:1994rz}.} solutions of the metric perturbation $\Phi$ and
the field fluctuation $\delta\phi_i$ in the longitudinal gauge and on the
super-horizon scales are given by
\begin{align}
\label{SRPhisol}
\Phi = & \epsilon C_1 + \frac{\sum_{i<j} C_{ij} \left( {V_i'}^2V_j - {V_j'}^2V_i \right)}
{3V^2} \, ,
\\
\label{SRdeltaphiisol}
\frac{\delta\phi_i}{\dot\phi_i} = & (1-\epsilon)\frac{C_1}{H} - 2H \frac{\sum_j
C_{ji}V_j}{V} \, ,
\end{align}
where $\epsilon$ is given by Eq.~(\ref{SRepsilon}). Here, $C_1$ and $d_i$ are
constants determined at the moment of horizon crossing given by
\begin{align}
\label{coeff_C1}
C_1 = & \left. \frac{-1}{\mpl^2} \sum_i \frac{V_i}{V_i'}\delta\phi_i \right|_\star
\, ,
\\
\label{coeff_di}
d_i = & \left. -\frac{3\delta\phi_i}{2V_i'} \right|_\star \, ,
\end{align}
with the subscript $\star$ denoting that both $C_1$ and $d_i$ are evaluated at the
moment of horizon crossing of the mode with momentum $k = |\mathbf{k}|$ and $C_{ij}
\equiv d_i - d_j$. Note that while $C_1$ is dimensionless, $d_i$ and thus $C_{ij}$
have mass dimension $-2$. We will denote the parts depending on $C_1$ as the
`adiabatic' modes, while those on $d_i$ as the `non-adiabatic' modes. Below, we will
denote them by the superscripts `(ad)' and `(nad)', respectively. Accordingly, the
energy density perturbation $\delta\rho_i$ can be written separately as
\begin{align}
\delta\rho_i = & \dot\phi_i \left[ \dot{\delta\phi_i^\ad} - \dot\phi_i\Phi^\ad
\right] + V_i'\delta\phi_i^\ad + \dot\phi_i \left[ \dot{\delta\phi_i^\nad} -
\dot\phi_i\Phi^\nad \right] + V_i'\delta\phi_i^\nad
\nonumber\\
\equiv & \delta\rho_i^\ad + \delta\rho_i^\nad \, ,
\end{align}
where $V_i' = dV_i/d\phi_i$. Then, using
\begin{equation}
\frac{d}{dt} \left[ \frac{\delta\phi_i^\ad}{\dot\phi_i} \right] = \Phi^\ad \, ,
\end{equation}
we obtain the adiabatic part of the curvature perturbation as
\begin{equation}\label{adia_relation}
\zeta^\ad \equiv \Phi^\ad + H \frac{\delta\rho_i^\ad}{\dot\rho_i} = \Phi^\ad + H
\frac{\delta\phi_i^\ad}{\dot\phi_i} = C_1 \, ,
\end{equation}
which is indeed constant. From this we can recover the well known ``adiabatic''
result
\begin{equation}
C_1 = \mathcal{R}_c \, ,
\end{equation}
with $\mathcal{R}_c$ being the comoving curvature perturbation. The non-adiabatic
part, the source of the time dependence, is from Eqs.~(\ref{SRPhisol}) and
(\ref{SRdeltaphiisol}),
\begin{align}\label{nadzeta}
\zeta_i^\nad = & - \frac{\sum_{j<k} C_{jk} \left( {V_j'}^2V_k - {V_k'}^2V_j
\right)}{3V^2} + 2H^2 \frac{\sum_j C_{ji}V_j}{V} \, .
\end{align}
With these adiabatic and non-adiabatic parts, the gauge invariant curvature
perturbations can be written as
\begin{align}
\label{PSzetai}
\zeta_i = & -C_1 - \Phi^\nad - H\frac{\delta\rho_i^\nad}{\dot\rho_i} \equiv -C_1 +
\zeta_i^\nad \, ,
\\
\label{PSzetatot}
\zeta = & -C_1 + \sum_i \frac{\dot\rho_i}{\dot\rho} \zeta_i^\nad \, .
\end{align}
It is very important to notice that {\em both} the adiabatic and non-adiabatic
solutions of $\Phi$ and $\delta\phi_i$ are contributing to the curvature
perturbation $\zeta$. More importantly, the evolution of $\zeta_i$ is purely due to
the non-adiabatic contribution. At this point, it would be instructive to illustrate
the origin of the name `adiabatic' and `non-adiabatic' modes: indeed, defining the
intrinsic isocurvature perturbation as
\begin{align}
\mathcal{S}_i \equiv & H \left( \frac{\delta{p}_i}{\dot{p}_i} -
\frac{\delta\rho_i}{\dot\rho_i} \right)
\nonumber\\
= & -\frac{2}{3}H \left[ -2\epsilon H \left( \frac{\sum_i d_iV_i}{V} - d_i \right) +
\frac{d}{dt} \left( \frac{\sum_i d_iV_i}{V} \right) \right] \, ,
\end{align}
we find that the adiabatic modes of $\Phi$ and $\delta\phi_i$ do not contribute to
$\mathcal{S}_i$. Also, the isocurvature perturbation between two fields is given by,
using Eq.~(\ref{isocurvatureS}),
\begin{equation}
S_{ij} = \frac{d}{dt} \left( \frac{\delta\phi_i^\nad}{\dot\phi} -
\frac{\delta\phi_j^\nad}{\dot\phi_j} \right) - 3H \left( \frac{\delta\phi_i^\nad}{\dot\phi_i}
- \frac{\delta\phi_j^\nad}{\dot\phi_j} \right) \, ,
\end{equation}
where the adiabatic modes are, again, canceled each other.

\subsubsection{During oscillation and afterward}

After inflation ends, the inflaton fields, in fact some of them already, enter the
oscillatory phase near the minimum of the effective potential. Ignoring the
pre-existing matter and radiation, the final curvature perturbations of matter and
radiation after all the inflatons decay can be written in terms of the curvature
perturbations of the scalar fields during the oscillatory period
as~\cite{Choi:2007fya}
\begin{align}
\label{zeta_m}
\zeta_m^\fin = & \sum_i s_i \zeta_i^\osc \,=-C_1 +  \sum_i s_i \zeta_i^{\nad\osc} \, ,
\\
\label{zeta_g}
\zeta_\gamma^\fin = & \sum_i r_i \zeta_i^\osc \,=-C_1 +  \sum_i r_i \zeta_i^{\nad\osc} \, ,
\end{align}
respectively. The coefficients $s_i$ and $r_i$ are given by~\cite{Choi:2007fya}
\begin{align}
\label{s_i}
s_i = & \frac{\Gmi}{\Gi} \Omega_i^\osc \left[ \sum_j
\frac{\Gamma_m^{(j)}}{\Gamma^{(j)}} \Omega_j^\osc \right]^{-1} \, ,
\\
\label{r_i}
r_i = & \prod_{j=i+1}^n \left( 1 -
\frac{3\rho_{\gamma{j}}/\rho_{\gamma{1}}}{4\sum_{k=1}^{j-1}
\rho_{\gamma{k}}/\rho_{\gamma{1}} + 3\rho_{\gamma{j}}/\rho_{\gamma{1}}} \right)
\frac{3\rho_{\gamma{i}}/\rho_{\gamma{1}}}{4\sum_{l=1}^{i-1}
\rho_{\gamma{l}}/\rho_{\gamma{1}} + 3\rho_{\gamma{i}}/\rho_{\gamma{1}}} \, .
\end{align}
We note that $\sum_i r_i = \sum_i s_i = 1$ by definition. Here, $\Omega_i^\osc$ is
the density parameter of the field $\phi_i$ at the beginning of the oscillatory
phase, and $\rho_{\gamma i}$ is the radiation energy density generated from the
decay of each $\phi_i$. Then the isocurvature perturbation between matter and
radiation after inflation is, from Eqs.~(\ref{isocurvatureS}), (\ref{zeta_m}) and
(\ref{zeta_g}),
\begin{equation}\label{Smr}
S_{m\gamma}^\fin = 3 \left[ \zeta_m^\fin - \zeta_\gamma^\fin \right]
= 3 \sum_i (s_i - r_i) \zeta_i^{\nad\osc} \, .
\end{equation}

\begin{figure}[h]
\centering
\rotatebox{90}{%
  \includegraphics[width=8cm]{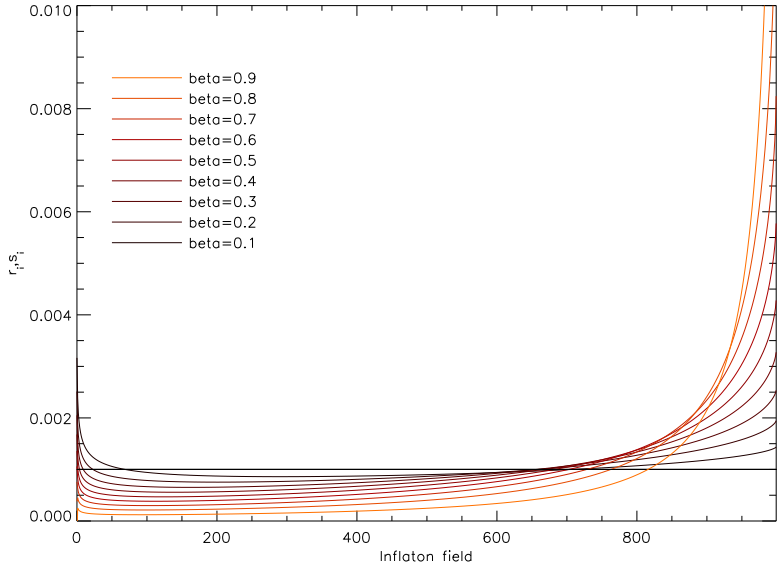}
}%
\caption{ The coefficients $s_i$ (black) and $r_i$ (red) used in the previous
section, calculated by Eqs.~(\ref{s_i_approx}) and (\ref{r_i_approx}) respectively
for the Mar{\v c}enko-Pastur distribution with different $\beta$ values. As we have
set the fraction of the matter decay rate the same,
$\Gamma_m^{(i)}/\Gamma_\gamma^{(i)}=10^{-6}$, $s_i$ is just a constant equal to
$N_\mathrm{fields}^{-1} = 10^{-3}$. Meanwhile, it is clear that the lighter fields
contribute more to the resulting radiation compared to heavier ones.
}%
\label{fig:coeffs}
\end{figure}

We can further simplify the above expressions in the presence of a large number of
inflaton fields, $N_\mathrm{fields}\gg1$, with not too different masses. In this
case, at the beginning of the oscillation phase (or equivalently the end of
inflation) each inflaton field contributes to the energy density with a similar
order of magnitude, so we can approximate
\begin{equation}\label{oscOmega}
\Omega_i^\osc \sim \frac{1}{N_\mathrm{fields}} \, .
\end{equation}
Using the fact that during the oscillatory phase the universe is practically matter
dominated, i.e. $a \propto t^{2/3}$ so that we can set the initial time as $t^\osc =
2/(3H_\osc)$, then we find~\cite{Choi:2007fya}
\begin{align}
\frac{\rho_{\gamma{i}}}{\rho_{\gamma{1}}} = &
\frac{\Gamma_\gamma^{(i)}\Omega_i^\osc/\Gamma^{(i)}}{\Gamma_\gamma^{(1)}\Omega_1^\osc/\Gamma^{(1)}}
\frac{[3/(2\Gamma^{(i)})]^{2/3}H_\osc^{1/6}}{[3/(2\Gamma^{(1)})]^{2/3}H_\osc^{1/6}}
\sim  \frac{\Gamma_\gamma^{(i)}}{{\Gamma^{(i)}}^{5/3}}
\frac{{\Gamma^{(1)}}^{5/3}}{\Gamma_\gamma^{(1)}} \, ,
\end{align}
where we have used Eq.~(\ref{oscOmega}). Thus, coefficients $s_i$ and $r_i$ are
simplified to
\begin{align}
\label{s_i_approx}
s_i \sim & \frac{\Gmi}{\Gi} \left[ \sum_j \frac{\Gamma_m^{(j)}}{\Gamma^{(j)}}
\right]^{-1} \, ,
\\
\label{r_i_approx}
r_i \sim & \prod_{j=i+1}^n \left[ 1 -
\frac{3\Gamma_\gamma^{(j)}/{\Gamma^{(j)}}^{5/3}}{4\sum_{k=1}^{j-1}
\Gamma_\gamma^{(k)}/{\Gamma^{(k)}}^{5/3} +
3\Gamma_\gamma^{(j)}/{\Gamma^{(j)}}^{5/3}} \right]
\frac{3\Gamma_\gamma^{(i)}/{\Gamma^{(i)}}^{5/3}}{4\sum_{l=1}^{i-1}
\Gamma_\gamma^{(l)}/{\Gamma^{(l)}}^{5/3} +
3\Gamma_\gamma^{(i)}/{\Gamma^{(i)}}^{5/3}} \, .
\end{align}
In Figure~\ref{fig:coeffs}, we present the coefficients $s_i$ and $r_i$ used in the
numerical calculations of the previous section.

Finally we make an estimation on $\zeta_i^{\nad\osc}$. To estimate the order of
magnitude of $\zeta_i^{\nad\osc}$, we first note that near the end of inflation $V_i
\sim V_j$ and $V_i' \sim V_j'$, as many fields are very close to the minimum. Thus,
from Eq.~(\ref{nadzeta}), the first term is comparably neglected and we can estimate
$\zeta_i^{\nad\osc}$ as
\begin{equation}\label{zeta_nad_osc}
\zeta_i^{\nad\osc}
\sim  \frac{2H_\fininf^2}{N_\mathrm{fields}}\sum_j C_{ji}
\equiv \sum_j \beta_{ij}d_j \, ,
\end{equation}
where the subscript `(end)' denotes the end of inflation. Also  we have made  a wild
guess $V_i \sim V/N_\mathrm{fields}$, i.e. each potential energy of the field
$\phi_i$ at the end of the inflation contributes equally to the total potential
energy. In the above we have defined $\beta_{ij}$ as the coefficients of the {\em
stochastic} variable $d_j$ and it is given by
\begin{equation}\label{betacoeff}
\beta_{ij} \sim \left\{
 \begin{split}
 & \frac{2H_\fininf^2}{N_\mathrm{fields}} & \textrm{for}\quad  j \neq i \, ,
 \\
 & \frac{2H_\fininf^2}{N_\mathrm{fields}} \left( 1 - N_\mathrm{fields} \right) &
 \textrm{for} \quad j = i \, .
 \end{split}
\right.
\end{equation}
Note that for the specific case of the potential given by Eq.~(\ref{multichaotic}),
it is known~\cite{Gong:2006zp} that $H_\fininf \approx
\sqrt{2}m_{N_\mathrm{fields}}/3$ with $m_{N_\mathrm{fields}}$ corresponding to the
lightest inflaton mass. Thus, we can write Eqs.~(\ref{zeta_m}) and (\ref{zeta_g})
simply as
\begin{align}
\label{zeta_m2}
\zeta_m^\fin = & -C_1 + \sum_{i,j} s_j \beta_{ji}d_i \, ,
\\
\label{zeta_g2}
\zeta_\gamma^\fin = & -C_1 + \sum_{i,j} r_j \beta_{ji}d_i \, ,
\end{align}
respectively. Note that all the coefficients $C_1$, $d_i$ and $\beta_{ij}$ are {\em
directly} related to the character of the inflaton fluctuations at the moment of
horizon crossing, and $s_i$ and $r_i$ to that of the decay of the inflaton fields.
This allows us to connect the initial conditions and/or parameters of the model with
the observable cosmological predictions with a reasonable accuracy (see
Table~\ref{table:summary}).

Now we can write the `final' curvature perturbation  as well as the isocurvature
perturbation between matter and radiation. After all of the inflaton fields decay
out, there remain only radiation and matter components in the universe. Radiation
and pressureless matter are perfect fluids with the equations of state of the form
$p = p(\rho)$, and the curvature perturbation of each species does not change.
However the total curvature perturbation does change depending on the energy density
ratio. They are expressed as 
\begin{align}
\label{zetafinal}%
\zeta^\fin = & \frac{4\rho_\gamma\zeta_\gamma^\fin +
3\rho_m\zeta_m^\fin}{4\rho_\gamma + 3\rho_m} \sim -C_1 + \sum_{i,j} \left[
\frac{4\rho_\gamma r_j + 3\rho_m s_j}{4\rho_\gamma + 3\rho_m} \beta_{ji}d_i \right]
\, ,
\\
S_{m\gamma}^\fin = & 3 \left[ \zeta_m^\fin - \zeta_\gamma^\fin \right] \sim 3
\sum_{i,j} (s_j-r_j) \beta_{ji}d_i \, .
\end{align}
We have tested the analytic estimations on $\zeta_m^\fin$, $\zeta_\gamma^\fin$ and
$S_{m\gamma}^\fin$, given by Eqs.~(\ref{zeta_m2}), (\ref{zeta_g2}) and (\ref{Smr})
respectively, against the full numerical results done in Section~\ref{numerical}. We
show the results in Table~\ref{table:summary}. We find that the analytic estimations
we have derived in this section predict the final curvature perturbations of both
matter and radiation within about $2\%$ accuracy. Note that the accuracy of the
analytic estimations changes with the number of the inflaton fields: it is improved
as the number of the inflaton fields increases. While the analytic estimation
captures the smallness of the isocurvature perturbation, it works within a factor of
$\mathcal{O}(1)$.

\subsubsection{Power spectra and spectral indices}

Now we can explicitly calculate the power spectra. The Fourier component of the
field fluctuation is obtained by using
\begin{equation}\label{fluctuation}
\delta\phi_i(\mathbf{k}) = \frac{H_\star}{\sqrt{2k^3}} e_i(\mathbf{k}) \, ,
\end{equation}
where $e_i(\mathbf{k})$ is a Gaussian random variable which satisfies
\begin{align}
\langle e_i(\mathbf{k}) \rangle = & 0 \, ,
\\
\langle e_i(\mathbf{k})e_j^*(\mathbf{q}) \rangle = &
\delta_{ij}\delta^{(3)}(\mathbf{k-q}) \, .
\end{align}
Then, substituting Eq.~(\ref{fluctuation}) into Eqs.~(\ref{coeff_C1}) and
(\ref{coeff_di}), we can write the final curvature and isocurvature perturbations in
momentum space as
\begin{align}
\label{zetafinal_fourier}
\zeta^\fin(k) = & \frac{H_\star}{\sqrt{2k^3}} \sum_i
\left( \frac{V_{i,\star}}{\mpl^2V_{i,\star}'}
- \frac{3}{2} \sum_j \frac{4\rho_\gamma r_j + 3\rho_m s_j}{4\rho_\gamma + 3\rho_m}
\frac{\beta_{ji}}{V_{i,\star}'} \right) e_i(\mathbf{k})
\nonumber\\
\equiv & \frac{H_\star}{\sqrt{2k^3}} \sum_i \left(
\frac{V_{i,\star}}{\mpl^2V_{i,\star}'} - p_i \right) e_i(\mathbf{k}) \, ,
\\
\label{Smr_momentum}
S_{m\gamma}^\fin(k) = & \frac{9}{2} \frac{H_\star}{\sqrt{2k^3}} \sum_i
\left[ \sum_j (r_j - s_j) \frac{\beta_{ji}}{V_{i,\star}'} \right] e_i(\mathbf{k}) \, .
\end{align}
Note that $p_i$, which encodes the post-inflationary evolution, has mass dimension
$-1$. Then, we define the power spectrum of a quantity $A_k$ as
\begin{equation}
\mathcal{P}_A \equiv \frac{k^3}{2\pi^2} \langle A_kA_{k} \rangle \, ,
\end{equation}
from which it is clear that the power spectrum of the correlation between $A_k$ and
$B_k$ should be given by $\langle A_kB_{k} \rangle$. Then, we can simply write the
power spectra of the final curvature perturbation, the matter-radiation isocurvature
perturbation and the correlation between them as
\begin{align}
 \label{Pcurv}
 \mathcal{P}_\mathrm{curv} = & \frac{V_\star}{12\pi^2\mpl^2} \sum_i
 \left( \frac{V_{i,\star}}{\mpl^2V_{i,\star}'} - p_i \right)^2 \, ,
 \\
 \label{Piso}
 \mathcal{P}_\mathrm{iso} = & \frac{27V_\star}{16\pi^2\mpl^2} \sum_i
 \left[ \sum_j (r_j - s_j) \frac{\beta_{ji}}{V_{i,\star}'} \right]^2 \, ,
 \\
 \label{Pcorr}
 \mathcal{P}_\mathrm{corr} = & \frac{3V_\star}{8\pi^2\mpl^2}
                               \sum_{i,j} \left(
                               \frac{V_{i,\star}}{\mpl^2V_{i,\star}'} - p_i \right)
                               \left( r_j - s_j \right)
                               \frac{\beta_{ji}}{V_{i,\star}'} \, ,
\end{align}
respectively. Note that it is clear for the single field case that we have $s_1 =
r_1 = 1$ and $p_1 = 0$, thus from Eq.~(\ref{Smr_momentum}) $S_{m\gamma} = 0$, i.e.
there exists no matter-radiation isocurvature perturbation. This is physically
reasonable: as the origin of both matter and radiation is the same, the curvature
perturbation of the inflaton component $\zeta_\phi$ is transferred to both $\zeta_m$
and $\zeta_\gamma$. Also note that $\mathcal{P}_\mathrm{corr}$ can have any sign,
depending on whether $\zeta^\fin$ and $S_{m\gamma}^\fin$ are correlated or
anti-correlated.

With the power spectra given by Eqs.~(\ref{Pcurv}) and (\ref{Piso}), we can easily
find the corresponding spectral indices. As everything is evaluated at the moment of
horizon crossing where $k = aH$, we have\footnote{Below, we omit the subscript
$\star$ to avoid too messy notations.}
\begin{equation}
d\log{k} \approx H dt \, .
\end{equation}
Thus, by defining
\begin{equation}
n - 1 \equiv \frac{d\log\mathcal{P}}{d\log{k}} \, ,
\end{equation}
we can write the index of the curvature power spectrum as
\begin{align}
n_\mathrm{curv} - 1 = & -2\epsilon - 2 \left[ \sum_j \left( \frac{V_j}{\mpl^2V_j'} -
p_j \right)^2 \right]^{-1} \sum_i \left( \frac{V_i}{\mpl^2V_i'} - p_i \right) \left[
\frac{V_i'}{V} \left( 1 - \frac{V_iV_i''}{{V_i'}^2} \right) + \frac{\dot{p_i}}{H}
\right] \, .
\end{align}
It is worthwhile to note that when the universe is completely dominated by
pressureless matter, the coefficient $p_i$ becomes very simple and so do the
spectral indices. In this case, from Eq.~(\ref{zetafinal_fourier}) we can easily
find
\begin{align}\label{pi_MD}
p_i = & \frac{3}{2V_i'} \sum_j s_j \beta_{ji} \, ,
\\
\frac{\dot{p_i}}{H} = & \mpl^2p_i\frac{V_i''}{V} \, ,
\end{align}
where we have assumed that both $s_i$ and $\beta_{ij}$ are independent on the moment
of horizon exit. Thus, the spectral index of the curvature power spectrum is
approximated as
\begin{align}\label{ncurv_estimate}
n_\mathrm{curv} - 1 = & -2\epsilon - \frac{2\mpl^2}{V} \sum_i V_i \left( 1 -
\frac{3}{2}\mpl^2 \frac{{\sum_j} s_j\beta_{ji}}{V_i} \right) \left[ 1 -
\frac{V_iV_i''}{{V_i'}^2} \left( 1 - \frac{3}{2}\mpl^2 \frac{\sum_k
s_k\beta_{ki}}{V_i} \right) \right]
\nonumber\\
& \hspace{3cm} \times \left[ \sum_l \left( \frac{V_l}{V_l'} \right)^2 \left( 1 -
\frac{3}{2}\mpl^2 \frac{\sum_m s_m\beta_{ml}}{V_l} \right)^2 \right]^{-1} \, .
\end{align}
Note that if we neglect $p_i$ and substitute Eq.~(\ref{pi_MD}) explicitly with
Eq.~(\ref{multichaotic}), we find after some calculations
\begin{equation}
n_\mathrm{curv} - 1 = -2\epsilon - \frac{4\mpl^2}{\sum_i\phi_i^2} \, ,
\end{equation}
which is consistent with the known results~\cite{Easther:2005zr,Gong:2006zp} where
the post-inflationary evolution is {\em neglected}. In the same way the spectral
index of the isocurvature power spectral is written as, once after inflation,
\begin{align}
n_\mathrm{iso} - 1 = & -2\epsilon + \frac{2\mpl^2}{V} \sum_i V_i'' \left[ \sum_j
(r_j - s_j) \frac{\beta_{ji}}{V_i'} \right]^2 \left\{ \sum_k \left[ \sum_l (r_l -
s_l) \frac{\beta_{lk}}{V_k'} \right]^2 \right\}^{-1} \, .
\end{align}
We can see that for small non-adiabatic contributions compared to the adiabatic one,
the curvature parts are mainly determined irrespective of the decay properties of
the inflatons. However the isocurvature perturbation is sensitive to them through
the coefficients $s_i$ and $r_i$. In Table~\ref{table:spectra}, we explicitly show
the values of the power spectra and the corresponding indices of the numerical
calculations we presented in the previous section. As can be read from the table,
$\mathcal{P}_\mathrm{iso}$ contributes about 0.1\% of $\mathcal{P}_\mathrm{curv}$.
Thus the normalization is mostly determined by the adiabatic contribution (see the
next section), and we have chosen the parameters to satisfy the current observations
on the order of magnitude basis. Also note that the indices $n_\mathrm{curv}$ and
$n_\mathrm{iso}$ are different.

\begin{table}[t!]
\centering
\begin{tabular}{c|cccccc}
\hline \hline
 & $\mathcal{P}_\mathrm{curv}$ & $n_\mathrm{curv}$ & $\mathcal{P}_\mathrm{iso}$ &
 $n_\mathrm{iso}$ & $\mathcal{P}_\mathrm{corr}$ & $n_\mathrm{corr}$
 \\
 \hline
 Fig.~\ref{case1} & $6.06920\times 10^{-9}$ & $0.958442$ & $5.32094\times 10^{-13}$ & $0.977479$  & $-2.38536\times 10^{-11}$ & $0.975164$
 \\
 \hline
 Fig.~\ref{case2} & $7.06238\times 10^{-9}$ & $0.958300$ & $1.86991\times 10^{-12}$ & $0.976821$  & $-4.47317\times 10^{-11}$ & $0.975055$
 \\
 \hline
 Fig.~\ref{case3} & $4.84587\times 10^{-9}$ & $0.952297$ & $1.71055\times 10^{-12}$ & $0.971534$  & $-4.76410\times 10^{-11}$ & $0.969105$
 \\
 \hline \hline
\end{tabular}
 \caption{
The values of the spectra and the corresponding indices of the numerical
calculations using the approximation of $\beta_{ij}$ in Eq.~(\ref{betacoeff}).}
 \label{table:spectra}
\end{table}

\section{Discussions}
\label{sec_discussions}

\subsection{$\delta{N}$ formalism}
\label{subsec_deltaN}

How is our result of the curvature perturbation presented in the previous section,
Eq.~(\ref{zetafinal}), or equivalently Eq.~(\ref{zetafinal_fourier}), related to the
one from the $\delta{N}$ formalism, for example presented in
Refs.~\cite{Easther:2005zr,Gong:2006zp}? With the potential $V = \sum_i
V_i(\phi_i)$, in the slow-roll approximation we can calculate the infinitesimal
$e$-folding as
\begin{equation}
dN = -\mpl^{-2} \sum_i \frac{V_i}{V_i'}d\phi_i \, ,
\end{equation}
so that the number of $e$-folds obtained {\em during} inflation is
\begin{equation}\label{efolds_during_inflation}
N = \int_{t_\ini}^{t_\fininf}Hdt = -\mpl^{-2} \sum_i \int_{\phi_i}^{\phi_{i\fininf}}
\frac{V_i}{V_i'}d\phi_i \, .
\end{equation}
If we specify the case $V = \sum_i m_i^2\phi_i^2/2$, we obtain
\begin{equation}\label{efold}
N = \frac{\sum_i \phi_i^2}{4\mpl^2} - \frac{\sum_i {\phi_{i\fininf}}^2}{4\mpl^2}
\equiv N_0 + F \, ,
\end{equation}
where for a large enough number of $\phi_i$ it is usually assumed that the slow-roll
phase is a good enough approximation until the end of inflation and that the
inflationary phase persists until very small value of $\phi_i$, so $\phi_{i\fininf}
\approx 0$. Thus the \textit{integration constant} $F$ is usually taken to be at
most of $\mathcal{O}(1)$ and can be neglected when we estimate $N$. However, $F$
does have an important implication: by specifying $F$, we can identify the point
where inflation ends, as can be read from Eq.~(\ref{efold}). Otherwise, we lose the
information on which trajectory the field has been evolving along. The reason is
that the end point completely specifies the {\em whole} evolution and tell us which
trajectory we choose, i.e. which universe we live in. That is, omitting $F$ is
equivalent to start from a minimum around which every direction is the same. This
can completely jeopardies the evaluation of the curvature perturbation which, as can
be read from Eq.~(\ref{efold})
\begin{equation}
\zeta = \delta{N} = \delta{N_0} + \sum_{i,j}
\frac{dF}{d\phi_{i\fininf}}\frac{d\phi_{i\fininf}}{d\phi_j}\delta\phi_j \, ,
\end{equation}
depends on the end point, or equivalently, on the trajectory: we may obtain an
abruptly large or small $\zeta$ and hence $\mathcal{P}_\zeta$ by choosing different
end points.

Now we return to the two-field case, and estimate the power spectrum taking into
account {\em only} the adiabatic component, i.e.
\begin{equation}
\zeta = -C_1 \, ,
\end{equation}
or in the momentum space
\begin{equation}
\zeta_k = \frac{H_\star}{2\sqrt{2k^3}\mpl^2} \left[ \phi_{1,\star} e_1(\mathbf{k}) +
\phi_{2,\star} e_2(\mathbf{k}) \right] \, .
\end{equation}
The power spectrum is then
\begin{equation}
\mathcal{P}_\mathrm{curv} = \frac{H_\star^2}{16\pi^2\mpl^4} \left( \phi_{1,\star}^2 +
\phi_{2,\star}^2 \right) \, .
\end{equation}
Meanwhile, taking {\em only} $N_0$ from Eq.~(\ref{efold}), $N_0 = (\phi_{1,\star}^2 +
\phi_{2,\star}^2)/4\mpl^2$ so that $N_{0,i} = \phi_{i,\star}/2\mpl^2$, thus
\begin{equation}\label{Pcurv_deltaN0}
\mathcal{P}_\mathrm{curv} = \left( \frac{H_\star}{2\pi} \right)^2 \sum_i N_{0,i}^2 =
\frac{H_\star^2}{16\pi^2\mpl^4} \left( \phi_{1,\star}^2 + \phi_{2,\star}^2 \right) \, ,
\end{equation}
which coincides with the result above. Thus, the adiabatic contribution is
associated with the leading contribution $N_0$ which does not take a specific
trajectory we follow into account. This is contained in $F$, associated with the
non-adiabatic contribution. That is, by taking only the `adiabatic' component of
$\zeta$, viz. $-C_1$, it reproduces the power spectrum which results from
$\delta{N_0}$. Omitting the information on the end point encoded in the other piece
which completely specifies the whole past evolution may completely spoil the result.
Fortunately, in this case the results with the integration coefficient $F$ is not
too different from the one without $F$, as there is only a single global minimum.
However if there are more than two equivalent vacua, omitting $F$ leads to vastly
different consequence. Further, in principle the total number of $e$-folds up to the
moment when there remains practically pressureless matter in the universe should be
written as
\begin{equation}
N = \int_{t_\ini}^{t_\fininf} H dt + \int_{t_\fininf}^{t_\mathrm{RH}} H dt +
\int_{t_\mathrm{RH}}^{t_\mathrm{eq}} H dt \, .
\end{equation}
To correctly predict the resulting curvature perturbation $\zeta$, we must take all
these pieces into account and perturb them. However,
Eq.~(\ref{efolds_during_inflation}) only corresponds to the first term and does not
care about the remaining two, which describe the post-inflationary evolution. Thus
it is clear by comparing Eqs.~(\ref{Pcurv}) and (\ref{Pcurv_deltaN0}) that, we need
to include the post-inflationary evolution as well as the integration constant $F$
for proper description of the curvature perturbation. In
Table~\ref{table:withandwithout}, we show the differences between the ``pure''
inflationary predictions on $\mathcal{P}_\mathrm{curv}$ and $n_\mathrm{curv}$ and
those including the contributions of the post-inflationary evolution. Since it is
obvious from Eqs.~(\ref{betacoeff}) and (\ref{zetafinal_fourier}) that during matter
domination there would be no difference with $s_i$ being the same for all $i$, we
boosted the difference by a factor $F_i = F_1 \exp\left[-(i-1)/s\right]$ with $s =
-999/\log{10^5}$, which makes an exponential distribution of $F_i$ such that
$F_{1000} = 10^5 F_1$. We set $F_1 = 0.001$.

\begin{table}[t!]
\centering
\begin{tabular}{c|cc}
\hline \hline
 & $\mathcal{P}_\mathrm{curv}$ & $n_\mathrm{curv}$
 \\
 \hline
 with $\beta_{ij}$ & $4.88637\times 10^{-9}$ & $0.952437$
 \\
 \hline
 without $\beta_{ij}$ & $4.90246\times 10^{-9}$ & $0.952492$
 \\
 \hline
 fractional difference & $3.29428\times 10^{-3}$ & $5.75981\times 10^{-5}$
 \\
 \hline \hline
\end{tabular}
 \caption{
The values of the curvature power spectra and the corresponding spectral indices
with and without the post-inflationary $\beta_{ij}$ terms. We have chosen the mass
distribution the uniform log one. The fractional changes are very small, and the
reason is discussed in Section~\ref{sec_largeiso}.}
 \label{table:withandwithout}
\end{table}

\subsection{Large isocurvature perturbation}
\label{sec_largeiso}

Now we turn to the possibility of large isocurvature perturbation between matter and
radiation $S_{m\gamma}$. It is well known that in the single field inflationary
model, it is impossible to generate $S_{m\gamma}$ unless we invoke additional
mechanism after inflation such as the curvaton
scenario~\cite{curvaton,isocurvature,Choi:2007fya}. In the present context,
$S_{m\gamma}$ arises because of the existence of a multiple number of the inflaton
fields. Thus observing $S_{m\gamma}$ itself or any of its possible consequences
amounts to seeing the effects of the multi-field inflation. From Eqs.~(\ref{zeta_m})
and (\ref{zeta_g}), while the adiabatic part of the solution is the same, only
non-adiabatic contribution can lead to different curvature perturbations associated
with matter and radiation. Thus, we can make the coefficients $s_i$ and $r_i$ very
different, and/or make $\zeta_i^{\nad\osc}$ very different.

First let us consider $\zeta_i^{\nad\osc}$. From Eq.~(\ref{coeff_di}), we can see
\begin{equation}
d_i \sim -{V_i'}^{-1} \sim -m_i^{-2} \, ,
\end{equation}
so that Eq.~(\ref{zeta_nad_osc}) is written as
\begin{equation}
\zeta_i^{\nad\osc} \sim -\frac{2H_\fininf^2}{N_\mathrm{fields}} \sum_j \left(
\frac{1}{m_j^2} - \frac{1}{m_i^2} \right) \, .
\end{equation}
We can immediately read two consequences. First, the contribution of
$\zeta_i^{\nad\osc}$ can be either positive or negative depending on the mass: for
the most (least) massive field, $\zeta_i^{\nad\osc}$ is maximally negative
(positive). This can be easily read from the numerical results of the previous
section, where the curvature perturbation associated with the lightest (heaviest)
field is larger (smaller) than the total one. Second, as inflation lasts longer and
longer, $\zeta_i^{\nad\osc}$ is driven towards zero as $H$ is, obvious from
Eq.~(\ref{BGeomH}), monotonically decreasing. This leads us to conclude that it
would not be easy to make $\zeta_i^{\nad\osc}$ very different from each other within
a separable potential with similar order of masses.

We can also try to make the coefficients $s_i$ and $r_i$ different for each field.
As noted in Ref.~\cite{Choi:2007fya}, there is hardly any model independent
prediction on the matter-isocurvature perturbation and the detail depends on the
specific decay rates. The decay rate we have used for the numerical computations is
essentially
\begin{equation}
\Gamma_\mathrm{\phi} \sim \frac{m_\phi^3}{\mpl^2} \, ,
\end{equation}
which is directly related to the mass of the field. While it would not be completely
impossible to contrive a non-trivial decay rate which can lead to large
matter-radiation isocurvature perturbation, it is not very likely that the fields
whose masses are spread over less than a decade (see Eq.~(\ref{MPdist}) for example)
have completely different decay rates to generate abruptly large matter-radiation
isocurvature perturbation. Especially, in our numerical computations in the previous
section we set all $s_i$ the same, and this leads to indeed negligible effects of
$S_{m\gamma}$ during matter dominated era. Therefore, most probably it would be very
hard to have large enough matter-radiation isocurvature perturbation in the context
of the current model.

Thus, we need to break the assumptions of the model we have explicitly investigated,
i.e. the potential includes interaction terms and is no more separable and/or the
decay rate is not constant so that $s_i$ spans a wide range of numbers as $r_i$ (see
Figure~\ref{fig:coeffs}). Further, it should be noticed that even if
$\mathcal{P}_\mathrm{iso}$ is negligible, the effects of the post-inflationary
evolution can still affect $n_\mathrm{curv}$ we can observe today: as can be read
from Eq.~(\ref{ncurv_estimate}), a non-trivial set of $s_i$ can make the difference
of $n_\mathrm{curv}$ an observationally detectable one from the estimate based on
naive inflationary predictions.

It is worthwhile to note again that while the formulae derived in this paper are
general and can be applied to any separable potential, the resulting small
isocurvature perturbation in our specific example is a peculiar property of the
investigated model, where the masses are distributed in the similar order of
magnitudes. If the difference of masses is large enough, such as double
inflation~\cite{doubleinflation}, then it is possible to obtain large isocurvature
perturbation.

We have also specifically assumed that the matter component is decoupled from
radiation already at the stage of generation. This can be possible only for the
extremely weakly interacting particles, such as gravitino or axino if they are
produced dominantly from the decay~\cite{gravitino}. For the thermally produced dark
matter, which is currently the popular candidate, the curvature perturbation of
matter is the same as that of radiation since they had common values during thermal
equilibrium. In this case there will be no or negligible isocurvature perturbation.

\section{Conclusions}
\label{sec_conclusions}

In this paper, we have studied the evolution of the curvature perturbation during
and after multi-field inflation. The inflaton fields $\phi_i$ decay into radiation
$\gamma$ and matter $m$ with the decay rates $\Gamma_\gamma^{(i)}$ and
$\Gamma_m^{(i)}$ respectively, which are assumed to be fixed by underlying physics.
We have presented the exact set of equations which describe the evolution of the
background quantities $\phi_i$, $\rho_\gamma$, $\rho_m$, $\rho$ and $H$,
Eqs.~(\ref{BGeomphii}), (\ref{BGeomrhog}), (\ref{BGeomrhom}), (\ref{BGcontinuity})
and (\ref{BGeomH}), respectively, as well as those for the perturbed quantities
$\delta\phi_i$, $\delta\rho_\gamma$, $\delta\rho_m$ and $\delta\rho$,
Eqs.~(\ref{eomdeltaphii}), (\ref{eomdeltarhog}), (\ref{eomdeltarhom}) and
(\ref{pcontinuity}). Using them, we have presented the curvature perturbations
associated with each component, $\zeta_\gamma$, $\zeta_m$ and $\zeta_i$ which
correspond to radiation, matter and $\phi_i$ respectively, Eqs.~(\ref{zetag}),
(\ref{zetam}) and (\ref{zetai}), as well as the total curvature perturbation $\zeta$
and the isocurvature perturbation between two components $S_{\alpha\beta}$,
Eqs.~(\ref{zetatot}) and (\ref{isocurvatureS}).

We have applied these set of equations to a particular simple model of multiple
chaotic inflation, with the potential given by Eq.~(\ref{multichaotic}). In
Figures~\ref{case1}, \ref{case2} and \ref{case3} we have present the numerical
results, and have found that $\zeta$ is continuously varying not only during
inflation but also afterward. We have presented an analytic argument to account for
the evolution of $\zeta$, which is solely due to the existence of the non-adiabatic
contributions of the curvature perturbation $\zeta_i^\nad$ given by
Eq.~(\ref{nadzeta}). $\zeta_i^\nad$ is responsible for the evolution of $\zeta$ as
well as the non-zero matter-isocurvature perturbation $S_{m\gamma}^\fin$,
Eq.~(\ref{Smr}). We have estimated $\zeta_m$, $\zeta_\gamma$ and $S_{m\gamma}$
analytically, and they are in reasonable agreement with the full numerical results
with $\mathcal{O}(1)$ factor, as shown in Table~\ref{table:summary}.

Throughout this study, we find several important points. It is clearly and
explicitly shown that the total curvature perturbation $\zeta$ is not fixed but is
constantly varying on the super-horizon scale, even after the end of inflation. This
indicates that any predictions of multi-field inflation based only on the
inflationary epoch may not capture important information about the evolution of the
universe: e.g. Eqs.~(\ref{Pcurv}) and (\ref{Pcurv_deltaN0}). This also implies that
until full radiation domination by the decay of the inflaton fields, the inflatons
themselves can modify $\zeta$ \`a la the curvaton mechanism. Finally, the
isocurvature perturbation between matter and radiation, which may be detected by
near future cosmological observations, becomes large enough only when the decay
rates are highly non-trivial. This suggests that large matter-radiation isocurvature
perturbation, if ever detected at all, will be a challenge to our current
understanding on the decay process of the inflaton fields.

\subsection*{Acknowledgments}

We thank Daniel Chung, Jai-chan Hwang and Misao Sasaki for helpful conversations.
 KYC is supported by the Ministerio de Educacion y Ciencia of Spain under Proyecto
Nacional FPA2006-05423 and by the Comunidad de Madrid under Proyecto HEPHACOS,
Ayudas de I+D S-0505/ESP-0346.
 JG is partly supported by the Korea Research Foundation Grant KRF-2007-357-C00014
funded by the Korean Government, and is currently supported in part by a VIDI and a
VICI Innovative Research Incentive Grant from the Netherlands Organisation for
Scientific Research (NWO).
 DJ acknowledges support from a Wendell Gordon Endowed Graduate Fellowship of
the University of Texas at Austin.

\end{document}